\documentclass[a4paper,11pt]{article}
\pdfoutput=1

\usepackage{jheppub}

\usepackage{subfig}
\usepackage{xspace}
\usepackage[countmax]{subfloat}
\usepackage{slashed}
\usepackage{longtable}

\usepackage{booktabs}
\usepackage{comment}
\usepackage{soul}

\usepackage{bbm}

\allowdisplaybreaks

\def\log{\text{log}}

\def\be{\begin{equation}}
\def\ee{\end{equation}}

\DeclareRobustCommand{\Sec}[1]{Sec.~\ref{#1}}

\DeclareRobustCommand{\Fig}[1]{Fig.~\ref{#1}}

\DeclareRobustCommand{\Eq}[1]{Eq.~(\ref{#1})}

\DeclareRobustCommand{\Ref}[1]{Ref.~\cite{#1}}
\DeclareRobustCommand{\Refs}[1]{Refs.~\cite{#1}}

\usepackage{color}
\definecolor{darkblue}{rgb}{0,0,0.5}
\definecolor{darkgreen}{rgb}{0,0.5,0}

\title{A Large-$N$ Expansion for Minimum Bias}

\author[a]{Andrew J.~Larkoski}
\affiliation[a]{Physics Department, Reed College, Portland, OR 97202, USA}

\author[b]{and Tom Melia}
\affiliation[b]{Kavli Institute for the Physics and Mathematics of the Universe (WPI), UTIAS, The University of Tokyo, Kashiwa, Chiba 277-8583, Japan}

\emailAdd{larkoski@reed.edu}
\emailAdd{tom.melia@ipmu.jp}

\abstract{
Despite being the overwhelming majority of events produced in hadron or heavy ion collisions, minimum bias events do not enjoy a robust first-principles theoretical description as their dynamics are dominated by low-energy quantum chromodynamics.  
In this paper, we present a novel expansion scheme of the cross section for minimum bias events that exploits an ergodic hypothesis for particles in the events and events in an ensemble of data.
We identify power counting rules and symmetries of minimum bias from which the form of the squared matrix element can be expanded in symmetric polynomials of the phase space coordinates.
This expansion is entirely defined in terms of observable quantities, in contrast to models of heavy ion collisions that rely on unmeasurable quantities like the number of nucleons participating in the collision, or tunes of parton shower parameters to describe the underlying event in proton collisions.
The expansion parameter that we identify from our power counting is the number of detected particles $N$ and as $N\to\infty$ the variance of the squared matrix element about its mean, constant value on phase space vanishes.
With this expansion, we show that the transverse momentum distribution of particles takes a universal form that only depends on a single parameter, has a fractional dispersion relation, and agrees with data in its realm of validity.
We show that the constraint of positivity of the squared matrix element requires that all azimuthal correlations vanish in the $N\to\infty$ limit at fixed center-of-mass energy, as observed in data.
The approach we follow allows for a unified treatment of small and large system collective behavior, being equally applicable to describe, e.g., elliptic flow in PbPb collisions and the ``ridge'' in pp collisions.
We also briefly comment on power counting and symmetries for minimum bias events in other collider environments and show that a possible ridge in $e^+e^-$ collisions is highly suppressed as a consequence of its symmetries.
}

\begin{document} 
\maketitle

%%%%%%%%%%%%%%%%%%%%%%%%%%%%%%%%%%
\section{Introduction}\label{sec:intro}
%%%%%%%%%%%%%%%%%%%%%%%%%%%%%%%%%%

At a high energy hadron or ionized nucleus collider, the dynamics of the collision is overwhelmingly dominated by quantum chromodynamics (QCD).  Because of the Landau pole and confinement of QCD at energies comparable to hadron masses, the momentum exchanged in these hadronic collisions is typically small compared to center-of-mass energies, leading to a production of a large number of observed particles with momenta not far above the QCD scale.  With the high luminosities of the experiments at the Large Hadron Collider (LHC) or the Relativistic Heavy Ion Collider (RHIC), it is not currently technically feasible to record all collision events for later analysis so triggers are employed that flag an event as interesting.  While passing triggers does introduce some bias in the events that are recorded, consideration of all events that pass the various triggers provides an ensemble of events that, as close as possible, probes low-energy QCD in a high-energy collision environment.  Developing an understanding of these so-called minimum bias events has been an active area of research for the past several decades \cite{Sjostrand:1987su,Durand:1988ax,Butterworth:1996zw,Borozan:2002fk,Ryskin:2009tj}.

Despite significant research efforts, a first-principles understanding of minimum bias events is lacking precisely because it exists at the scale for which QCD becomes strongly interacting.  Thus, minimum bias events are typically described by models with many parameters, like that used in modern general-purpose event simulation programs like Pythia \cite{Sjostrand:2006za} or Herwig \cite{Bahr:2008dy}.  A hydrodynamics limit is often used to model and interpret collisions of heavy ions which result in extremely high particle multiplicities  \cite{Kolb:2003dz,Romatschke:2009im,deSouza:2015ena},  providing evidence for production of the quark-gluon plasma (QGP) \cite{BraunMunzinger:2007zz,Shuryak:2008eq,Nouicer:2015jrf,Pasechnik:2016wkt}.  However, with a model, experimental results can only be interpreted within the context of that model and without a more general framework may obscure other consistent descriptions of data.  What is more, such a description relies on unphysical or unmeasurable parameters (like temperature or chemical potential), and properties of collisions are often described in terms of the unmeasurable impact parameter or overlap of the nuclei. In this paper our philosophy is to propose and explore a framework in which certain collider observables may be captured in a model independent fashion, in terms of purely physical/measurable quantities.

One instance where not being tied to a particular model(s) may be particularly useful is in the study of the relatively newly discovered phenomena of collective behavior in small systems such as pp collisions at the LHC and pA collisions at RHIC (see, e.g., \Ref{doi:10.1146/annurev-nucl-101916-123209} for a review). For instance, in order to disentangle evidence of the production of small `droplets' of QGP from other physics that leads to particle correlations, it would be desirable to have a framework that can interpolate between different system sizes, and collision energies.

The basic idea of the model independent approach we explore is to work directly with the measured particles, and  constrain the $S$-matrix of the minimum bias events. This philosophy  aligns with a bootstrap approach to QCD which is actively being pursued for low-multiplicity processes in quantum field theory \cite{Paulos:2016fap,Paulos:2016but,Paulos:2017fhb,Cordova:2018uop,Mazac:2018mdx,Mazac:2018ycv,Cordova:2019lot,Karateev:2019ymz,Correia:2020xtr, Homrich:2019cbt, Komatsu:2020sag,Caron-Huot:2020adz,Guerrieri:2021tak}. The nature and high-multiplicity of minimum bias events offers some simplifications by appealing to the principle of ergodicity: a given particle is representative of any particle in the event. Further, events with $N$ and $N+1$ particles have approximately equal descriptions, and thus we can expect to be able to expand in the multiplicity parameter $N$.
In the absence of a first-principles understanding, we look for a detector-level effective  description of minimum bias for which there exists concrete power counting and symmetries that guide the description and has parameters that can be fit to data that are then ultimately described in QCD. 

At a hadron collider, there are always particles that are unobservable as they go straight down the beampipe. We show that  integrating out these particles from the description of events has the effect of smearing the phase space of the observed particles in the detector in a universal fashion, reproducing the flat-in-(pseudo)rapidity distribution argued by Feynman~\cite{Feynman:1969ej}. The expansion of the matrix element of the observed particles then provides fluctuations about this flat distribution. 

Importantly, the approach is applicable to observables that are binned in $N$. That is, it does not provide a first-principles description of  fluctuations in multiplicity. Thus the kind of statements that can be made are, for example, how {\it normalized} distributions, binned in $N$, may change as a function of $N$. We also assume a fixed collider energy, $Q$, when taking the large $N$ limit, in the sense that we do not consider a `t Hooft coupling-like limit of taking $Q,N\to\infty$ at fixed $Q/N$; statements about the scaling with $Q$, however, are obtained.

Our aim is to explore concrete, quantitative predictions that must follow from consistency of the above-described approach to minimum bias.  We first define the power counting scheme for minimum bias events and identify its symmetries exclusively in terms of measurable quantities, just like the starting point of any effective field theory (EFT).  Through this procedure, the relevant expansion parameter we identify is novel and unlike familiar EFTs.  We formally take the number of observed particles $N\gg 1$, which is similar to a hydrodynamics limit.  From the general expression for the cross section for production of $N$ final state particles in minimum bias events, we expand the squared matrix element in symmetric polynomials of the phase space coordinates, ordered in powers of $1/N$.  Among predictions that we validate in collider data include:
\begin{itemize}

\item In the $N\to\infty$ \ limit the symmetries of minimum bias events and the central limit theorem require that the squared matrix element is exclusively a function of the squared total energy of the observed final state particles.

\item Inspired by the soft and collinear singularities of perturbative QCD, we show that the distribution of particle transverse momentum is universal and depends on a single parameter in the large-$N$ limit.  Further, the transverse momentum distribution implies that particles in minimum bias events have a fractional dispersion relation $\omega \propto k_\perp^{2/3}$.

\item By positivity of the squared matrix element, we demonstrate that all pairwise particle azimuthal correlations necessarily vanish as $N\to \infty$ at fixed center-of-mass collision energy. 

\item Long-distance pairwise particle pseudorapidity correlations are a consequence of factorization of the squared matrix element as $N\to\infty$.

\item Non-trivial pairwise particle azimuthal correlations are highly suppressed in minimum bias events in $e^+e^-$ collisions because of the full Lorentz invariance of the observed final state particles.

\end{itemize}
These results and others suggest that development of an effective theory for minimum bias would describe a wide range of phenomena and could be matched to perturbative QCD or hydrodynamics.

This paper is organized as follows.  In \Sec{sec:ppmin}, we establish the power counting for minimum bias in identical hadron or heavy ion collisions and the corresponding symmetries that those events enjoy.  We show how to construct the squared matrix element in terms of symmetric polynomials of the phase space coordinates.  In \Sec{sec:datacomp}, we use this effective form of the minimum bias cross section to interpret collider data.  We show that this large-$N$ expansion can provide quantitative and precise predictions that agree with data in its realm of applicability.  In \Sec{sec:other}, we describe the power counting and symmetries for different collider environments.  Because minimum bias events in $e^+e^-$ collisions enjoy extensive symmetry, we show that these symmetries strongly suppress pairwise particle azimuthal correlations in the large-$N$ limit.  In \Sec{sec:concs}, we summarize our findings and discuss future avenues to further develop an effective theory of minimum bias.

%%%%%%%%%%%%%%%%%%%%%%%%%%%%%%%%%%
\section{Power Counting and Symmetries of pp/AA Collision Min-Bias}\label{sec:ppmin}
%%%%%%%%%%%%%%%%%%%%%%%%%%%%%%%%%%

We first establish the power counting and symmetries we employ to describe minimum bias events.  This will correspondingly precisely define what we theoretically mean by ``minimum bias'', which is distinct from the experimental definition.  We will focus on minimum bias events at a hadron collider, in particular, proton-proton (pp) or identical heavy ion (AA) collisions.  We assume that the lab frame is also the center-of-mass frame of the collisions.  The center-of-mass collision energy $Q$ is assumed to be much larger than the masses of the initial, colliding particles.  Throughout this paper, we assume that $Q$ is an arbitrary, but fixed, energy, but will occasionally comment on the dependence of distributions of observables as a function of $Q$.  Additionally, we assume that the experimental detector cannot measure all particles and that there is an unmeasureable beam region.  Concretely, we assume that there is a maximum pseudorapidity $\eta_{\max}$ of the detector and particles with pseudorapidity $|\eta|>\eta_{\max}$ are lost down the beampipe.  Finally, we assume that our detector can only measure particle momenta, but has perfect angular and energy resolution.

\begin{figure}
\begin{center}
\includegraphics[width=14cm]{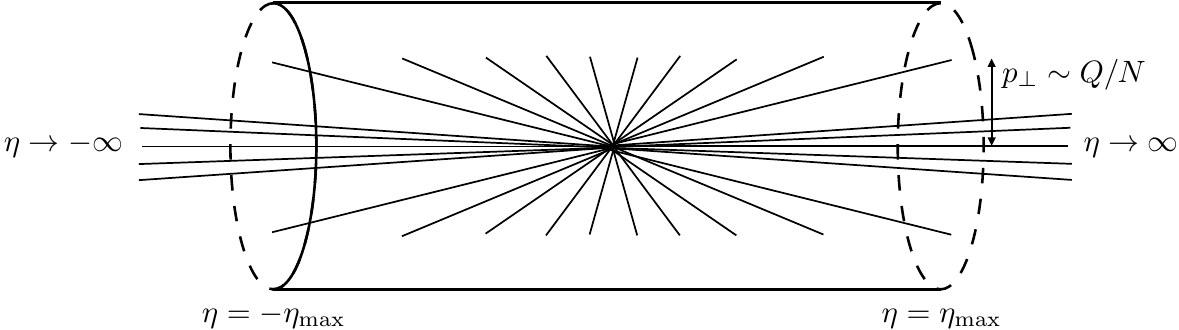}
\caption{Schematic illustration of detector (cylinder), its pseudorapidity range $\eta_{\max}$, the beam regions around $\eta = \pm\infty$, and the characteristic transverse momentum of detected particles.
\label{fig:detect}
}
\end{center}
\end{figure}

With these assumptions, we can establish power counting that defines minimum bias events and the corresponding symmetries of our collision experiment and of the detected final state particles.  The power counting that we take is as follows.
\begin{enumerate}

\item The beam is a small angular region outside the detection apparatus and we restrict our description of the event to far from the beam region, where detected particle pseudorapidity satisfies $|\eta|\sim 1\ll \eta_{\max}$.

\item  We assume that the mass of the particles is irrelevant and so detected particle transverse momentum $p_\perp$ is parametrically larger than the QCD scale or pion mass, $p_\perp \gg m_\pi$. 

\item The momentum lost down the beam region is an order-1 fraction of the center-of-mass energy $Q$. 

\item The number of detected particles $N$ for which their pseudorapidity $|\eta|\ll\eta_{\max}$ is large: $N\gg1$.

\item We assume that the mean transverse momentum of the detected particles is representative of all particles' momenta and so the mean and the root mean square momenta are comparable: $\langle p_\perp\rangle\sim \sqrt{\langle p_\perp^2\rangle}$.
\end{enumerate}
We will see below how these power counting assumptions then imply that our expansion parameter is $1/N \sim p_\perp/Q\ll 1$.\footnote{Actually, as we will see in the explicit examples below, there is a (dominant) additional fixed factor of $(Q/p_{\perp\text{cut}})^\frac{1}{2}$ in this scaling relation, with $p_{\perp\text{cut}}$ the experimental cut on transverse momentum.  This factor is fixed over an ensemble of events and we absorb it into `$\sim$' here and in the following section when evaluating the scaling in $1/N$ of terms in the expansion.} Additionally, these power counting assumptions imply that the minimum bias events we consider satisfy something like an ergodic hypothesis.  That is, we assume that any individual particle in an event is representative of all particles in an event.  Thus, averages over particles in an event are equivalent to averages of events over an ensemble.  This ergodic hypothesis will have important implications for the structure of the squared matrix element in the $N\to\infty$ limit.  An illustration of the physical configuration established by these power counting assumptions is provided in \Fig{fig:detect}.

Hadron collision events that satisfy these power counting requirements then enjoy the following symmetries:
\begin{enumerate}

\item O(2) rotation and reflection symmetry  about the beam,

\item reflection of the beam $\eta\to-\eta$ because of the identical colliding particles,

\item $S_N$ permutation symmetry in all $N$ detected particles, and

\item translation symmetry in pseudorapidity, $\eta\to \eta+\Delta \eta$.
\end{enumerate}
Most of these symmetries should be apparent as a consequence of the familiar cylindrical shape of particle detectors.  Because we assume only particle momenta are measured, no distinguishing information of particles (like electric charge) is collected, and hence there is complete permutation symmetry. By our power counting, we assume that the beam region is defined by $\eta_{\max}\gg1$.  Taking the formal limit $\eta_{\max}\to\infty$, a translation in $\eta$ by a finite amount $\Delta \eta$ can never move particles out of the detected region into the beam region, or vice-versa.  

We then define minimum bias events in hadron collisions as those that satisfy the power counting assumptions, and therefore inherit the symmetries.  Experimentally, minimum bias events are typically defined as those that pass a minimal trigger threshold for, say, activity in a forward calorimeter.  As the name suggests, passing a trigger does bias the event selection somewhat, but this bias can potentially be reduced by performing other cuts on these events.  For example, to force an event into the regime described by our power counting, one could require that a fixed fraction of particles have a transverse momentum within some range about the mean transverse momentum.  Or, one could require that an event does not have any identified jets with transverse momentum larger than some fraction of the center-of-mass energy.  We will return to jets below, and explicitly demonstrate that the existence of hard jets in the final state violates our power counting.

Because of the high luminosity of modern hadron colliders, truly zero bias events can collected by exploiting pile-up, or secondary hadron collisions per bunch crossing.  A hard hadron collision that produces jets or high-energy leptons, for example, would pass the triggers, but then secondary hadron collision events in the bunch crossing would have no requirements placed on them, and still be measured.  The challenge with extracting these zero bias events is then pushed to the ability to distinguish the point of collision where different particles were produced.  Any number of our power counting assumptions can be relaxed to provide a more realistic description of these events; however, we will find that even these strongly constraining assumptions will be able to explain and understand a wide breadth of data in various collider environments.
 
\subsection{Effective Form of the Cross Section}

We will use the power counting and symmetries to provide an effective description of the scattering cross section for these minimum bias events exclusively in terms of properties of the measured particles.  To derive this description, we start from the completely general formula for the cross section, considering $2\to N+N_{B_1}+N_{B_2}$ scattering events, where $N_{B_1}+N_{B_2}$ particles are beam remnants; i.e., $N_{B_1}$ particles have pseudorapidity $\eta > \eta_{\max}$, $N_{B_2}$ particles have $-\eta > \eta_{\max}$, and $N$ particles have $|\eta|< \eta_{\max}$.  The cross section $\sigma$ for this scattering can be expressed in generality as
\begin{align}
\sigma &\sim \int d\Pi_{N+N_{B_1}+N_{B_2}}\, |{\cal M}(1,\dotsc,N,N+1,\dotsc, N+N_{B_1},N+N_{B_1}+1,\dotsc,N+N_{B_1}+N_{B_2})|^2 \nonumber\\
&\sim  \int \prod_{i=1}^N\left[p_{\perp i}\, dp_{\perp i}\, d\eta_i\, \frac{d\phi_i}{2\pi}  \right]\, \prod_{j=1}^{N_{B_1}}\left[dk_j^+\, dk_j^-\, \frac{d\phi_j}{2\pi} \right]\, \prod_{k=1}^{N_{B_2}}\left[dk_k^+\, dk_k^-\, \frac{d\phi_k}{2\pi}  \right]\,
\nonumber\\
&
\hspace{0.5cm}\times|{\cal M}(1,\dotsc,N+N_{B_1}+N_{B_2})|^2\, \delta\left(
Q - \sum_{i=1}^N p_{\perp i}e^{\eta_i} - \sum_{j=1}^{N_{B_1}} k_j^-- \sum_{k=1}^{N_{B_2}} k_k^-
\right)\nonumber\\
&\hspace{0.5cm}
\times \delta\left(
Q - \sum_{i=1}^N p_{\perp i}e^{-\eta_i} - \sum_{j=1}^{N_{B_1}} k_j^+- \sum_{k=1}^{N_{B_2}} k_k^+
\right)\,\delta^{(2)}\left(
\sum_{i=1}^N \vec p_{\perp i}+\sum_{j=1}^{N_{B_1}} \vec p_{\perp j}+\sum_{k=1}^{N_{B_2}} \vec p_{\perp k}
\right)\,.
\end{align}
In writing this expression, we have ignored overall factors of the center-of-mass energy $Q$ and powers of $2\pi$.  In the first line, we implicitly write the cross section as the integral of the squared matrix element over $N+N_{B_1}+N_{B_2}$-body phase space, and then in the remaining lines, expand out the massless Lorentz-invariant integration measure and momentum-conserving $\delta$-functions.  $k^+_i$ and $k^-_i$ are the plus and minus components of particle $i$'s lightcone momentum with respect to the beam axis (the $z$ axis).  The momentum-conserving $\delta$-functions fix the lab frame to be the center-of-mass frame of the collisions.

Now, with the assumption that $\eta_{\max}\gg 1$, we will formally take $\eta_{\max} \to\infty$, so the beam remnants live at $\pm \infty$ pseudorapidity.  With that assumption, their transverse momenta are 0 and one component of their lightcone momenta is also 0, depending on whether the particles are in beam 1 or beam 2.  These assumptions simplify the momentum conserving $\delta$-functions and the cross section becomes  
\begin{align}
\sigma &\sim   \int \prod_{i=1}^N\left[p_{\perp i}\, dp_{\perp i}\, d\eta_i\, \frac{d\phi_i}{2\pi}  \right]\, \prod_{j=1}^{N_{B_1}}\left[dk_j^+\, dk_j^-\, \frac{d\phi_j}{2\pi} \right]\, \prod_{k=1}^{N_{B_2}}\left[dk_k^+\, dk_k^-\, \frac{d\phi_k}{2\pi}  \right]\,|{\cal M}(1,\dotsc, N+N_{B_1}+N_{B_2})|^2\nonumber \\
&
\hspace{0.5cm}\times \delta\left(
Q - \sum_{i=1}^N p_{\perp i}e^{\eta_i} - \sum_{j=1}^{N_{B_1}} k_j^-
\right)\, \delta\left(
Q - \sum_{i=1}^N p_{\perp i}e^{-\eta_i} - \sum_{k=1}^{N_{B_2}} k_k^+
\right)\,\delta^{(2)}\left(
\sum_{i=1}^N \vec p_{\perp i}
\right)\,.
\end{align}
It may seem like the assumption that the beam remnants carry no net transverse momentum is too constraining.  However, assuming that the number of detected particles $N$ is large and the correlations between the detected particles are relatively weak, the net transverse momentum of the detected particles is necessarily small.  The detected particles in an individual event can be imagined to randomly walk in the transverse momentum plane and the average net transverse momentum in an ensemble of events will be 0.  As a random walk, the standard deviation of the net transverse momentum in this ensemble of events scales like $1/\sqrt{N}$ at fixed collision energy $Q$.  Then, to leading approximation in the large-$N$ limit, we assume that the net transverse momentum of the detected particles is 0.  It would be interesting to determine the consequences of relaxing this assumption, but we leave that to future work.

Because we do not measure the particles in the beam region, we want to integrate over them.  With the assumption of permutation symmetry of the particles and the fact that we assume that all beam remnants travel in the exact same direction down either pipe, the only quantity that we can be sensitive to in experiment is their total momentum.  That is, the squared matrix element can only depend on a function of the total plus or minus lightcone momentum that goes down the beampipes:
\begin{align}
&|{\cal M}(1,\dotsc,N,N+1,\dotsc, N+N_{B_1},N_{B_2})|^2 \\
&
\hspace{1cm}\sim f\left(
 \sum_{j=1}^{N_{B_1}} k_j^-
,
\sum_{k=1}^{N_{B_2}} k_k^+
\right)\, \frac{|{\cal M}(1,\dotsc,N,N+1,\dotsc, N+N_{B_1},N_{B_2})|^2}{f\left(
 \sum_{j=1}^{N_{B_1}} k_j^-
,
\sum_{k=1}^{N_{B_2}} k_k^+
\right)}\nonumber\\
&\hspace{1cm}\equiv f\left(
 \sum_{j=1}^{N_{B_1}} k_j^-
,
\sum_{k=1}^{N_{B_2}} k_k^+
\right)\, |{\cal M}(1,2,\dotsc,N)|^2\nonumber\,.
\end{align}
Here, $f$ is some function of the total lightcone momentum that goes in either beam direction.  In the final line, we have expressed squared matrix element with dependence on the measured particles $1,\dotsc, N$ and dependence on the total lightcone momentum in the beam regions is left implicit.  We will return to the explicit form of this matrix element later. The $\eta\to -\eta$ symmetry of the event requires that the function $f$ is symmetric in its arguments: $f(x,y) = f(y,x)$.  Further, by the $\eta\to \eta+\Delta \eta$ translation symmetry, the argument of the function $f$ must be the product of the total plus and minus lightcone momenta, because lightcone momentum components transform homogeneously under boosts along the $z$ axis.  Then, the squared matrix element can be expressed as
\begin{align}
|{\cal M}(1,\dotsc,N,N+1,\dotsc, N+N_{B_1},N_{B_2})|^2 \sim f\left(
 \sum_{j=1}^{N_{B_1}} k_j^-
\sum_{k=1}^{N_{B_2}} k_k^+
\right)\, |{\cal M}(1,2,\dotsc,N)|^2\,.
\end{align}

By momentum conservation and choosing the collision frame to be the center-of-mass frame, we also know that the total lightcone momentum lost down the beam regions is related to the lightcone momentum of the detected particles.  In particular, 
\begin{align}
& \sum_{j=1}^{N_{B_1}} k_j^- = Q-\sum_{i=1}^N p_{\perp i}e^{\eta_i}\,, &\sum_{k=1}^{N_{B_2}} k_k^+ = Q-\sum_{i=1}^N p_{\perp i}e^{-\eta_i}\,.
\end{align}
Thus, we can equivalently express the function $f$ exclusively in terms of directly measurable quantities.  We define
\begin{align}
&k^-\equiv \sum_{i=1}^N p_{\perp i}e^{\eta_i}\,, &k^+\equiv\sum_{i=1}^N p_{\perp i}e^{-\eta_i}\,,
\end{align}
so that the argument of the function $f$ is just the product $k^+k^-$, by the same reasoning above.  That is, the matrix element can be expressed as
\begin{align}
|{\cal M}(1,\dotsc,N,N+1,\dotsc, N+N_{B_1},N_{B_2})|^2 \sim f\left(
k^+k^-
\right)\, |{\cal M}(1,2,\dotsc,N)|^2\,,
\end{align}
for some function $f(k^+k^-)$ and our power counting assumes that the momentum lost down the beam regions (or, the $z$-axis boost of the detected particles), is of a comparable size to the center-of-mass energy, $ Q^2$.  

Then, with the power counting and symmetries enforced on the form of the matrix element, in the cross section we can integrate over the momentum lost down the beams.  The cross section can then be expressed as
\begin{align}\label{eq:xsecmaster}
\sigma &\sim \int_0^Q dk^+ \int_0^Q dk^- \int \prod_{i=1}^N\left[p_{\perp i}\, dp_{\perp i}\, d\eta_i\, \frac{d\phi_i}{2\pi}  \right]\,\prod_{j=1}^{N_{B_1}}\left[dk_j^+\, dk_j^-\, \frac{d\phi_j}{2\pi} \right]\, \prod_{k=1}^{N_{B_2}}\left[dk_k^+\, dk_k^-\, \frac{d\phi_k}{2\pi}  \right]\,f\left(
k^+k^-
\right)\nonumber \\
&
\hspace{1cm}\times  |{\cal M}(1,2,\dotsc,N)|^2\, \delta\left(
Q-k^- -\sum_{j=1}^{N_{B_1}} k_j^-
\right)\, \delta\left(
Q-k^+ - \sum_{k=1}^{N_{B_2}} k_k^+
\right)\,\delta^{(2)}\left(
\sum_{i=1}^N \vec p_{\perp i}
\right)
\nonumber\\
&\hspace{1cm}
\times \delta\left(
k^--\sum_{i=1}^N p_{\perp i}e^{\eta_i}
\right)\,\delta\left(
k^+-\sum_{i=1}^N p_{\perp i}e^{-\eta_i}
\right)\nonumber
\\
&\sim  \int_0^Q dk^+ \int_0^Q dk^- \int \prod_{i=1}^N\left[p_{\perp i}\, dp_{\perp i}\, d\eta_i\, \frac{d\phi_i}{2\pi}  \right]\, f\left(
k^+k^-
\right)\, |{\cal M}(1,2,\dotsc,N)|^2 \nonumber\\
&
\hspace{1cm}\times \delta\left(
k^- - \sum_{i=1}^N p_{\perp i}e^{\eta_i}
\right)\, \delta\left(
k^+ - \sum_{i=1}^N p_{\perp i}e^{-\eta_i}
\right)\,\delta^{(2)}\left(
\sum_{i=1}^N \vec p_{\perp i}
\right)\,.
\end{align}
In doing the integrals over the unobservable beam remnant particles, we have ignored overall factors of the center-of-mass energy $Q$, and additional dependence on the product $k^+k^-$ has been absorbed in the definition of $f(k^+k^-)$. Because $f(k^+k^-)$ is a physical squared matrix element, we make the reasonable assumption that it is finite and analytic on its domain.  It can therefore be expanded in an appropriate basis of orthonormal polynomials, and this sum can be truncated for approximation to fit data.

In analysis that we will present later, it will be useful to know the volume of this $N$-body phase space smeared by a boost along the beam axis.  For a center-of-mass energy $Q$, the volume of $N$-body phase space is
\begin{equation}\label{eq:psvol}
\int d\Pi_N = (2\pi)^{4-3N}Q^{2N-4}\frac{2\pi^{N-1}}{(N-1)!(N-2)!}\,.
\end{equation}
Setting the function $f(k^+k^-)$ and the squared matrix element $|{\cal M}(1,2,\dotsc,N)|^2$ to unity, the volume of this smeared phase space is therefore
\begin{align}
&\int_0^Q dk^+ \int_0^Q dk^- \int \prod_{i=1}^N\left[p_{\perp i}\, dp_{\perp i}\, d\eta_i\, \frac{d\phi_i}{2\pi}  \right]\,\\
&
\hspace{2cm}\times \delta\left(
k^- - \sum_{i=1}^N p_{\perp i}e^{\eta_i}
\right)\, \delta\left(
k^+ - \sum_{i=1}^N p_{\perp i}e^{-\eta_i}
\right)\,\delta^{(2)}\left(
\sum_{i=1}^N \vec p_{\perp i}
\right)\nonumber\\
&\hspace{1cm}=(2\pi)^{4-3N}\frac{2\pi^{N-1}}{(N-1)!(N-2)!}\int_0^Q dk^+ \int_0^Q dk^-\, (k^+k^-)^{N-2}\nonumber\\
&\hspace{1cm}=(2\pi)^{4-3N}Q^{2N-2}\frac{2\pi^{N-1}}{(N-1)^2(N-1)!(N-2)!}\nonumber\,.
\end{align}
We also note that the topology of the smeared phase space is that of a $(3N-2)$-ball, found by integrating over two of the dimensions of the $N$-body phase space manifold \cite{Larkoski:2020thc}.

\subsection{Expansion of the Squared Matrix Element}

Having established the form of the cross section for minimum bias events expressed exclusively in terms of observable particle momenta, we now use the power counting and symmetries to expand the squared matrix element $|{\cal M}(1,2,\dotsc,N)|^2$.  The natural expansion parameter in which to do this according to our power counting is the number of detected particles $N\gg 1$.  Namely, we would want to determine the momentum dependence of the squared matrix element systematically at every power of $1/N$, with undetermined parameters that can be fit to data.  We will do this in a few steps.  First, we will present an expansion of the squared matrix element in inverse powers of the center-of-mass energy $Q$, for which terms can be established order-by-order through use of momentum conservation.  From this expansion, we can identify the scaling with $N$ for each of the terms and reorganize the expansion according to our power counting rules.

We first note that, with this construction, the leading approximation of the squared matrix element is that it is constant.  As we can fix overall normalization later, the leading approximation of squared matrix element in the $1/Q$ expansion is
\begin{equation}
|{\cal M}(1,2,\dotsc,N)|^2 = 1 + {\cal O}(Q^{-1})\,.
\end{equation}
A linearly-independent set of momentum-dependent terms at higher powers in $1/Q$ can be established using momentum conservation.  The transverse and longitudinal momentum conservation equations of the detected particles are:
\begin{align}\label{eq:momconserv}
0&=\left(
\sum_{i=1}^N \vec p_{\perp i}
\right)^2 = \sum_{i=1}^N p_{\perp i}^2 + \sum_{i\neq j}^N p_{\perp i}p_{\perp j}\cos(\phi_i-\phi_j)\,,\\
k^+k^-&=\left(
\sum_{i=1}^N p_{\perp i}e^{-\eta_i}
\right)\left(
\sum_{j=1}^N p_{\perp j}e^{\eta_j}
\right) = \sum_{i=1}^N p_{\perp i}^2 + \sum_{i\neq j}^N p_{\perp i}p_{\perp j}\cosh(\eta_i-\eta_j)\,.\nonumber
\end{align}
On the right, we have expanded the equalities in terms of permutation-symmetric polynomials of the phase space coordinates.  The symmetries of minimum bias events forbid terms at order $1/Q$ (and actually any odd power of this) and these fundamental momentum conservation equations define a linearly-independent set of terms at order $1/Q^2$.  We can then write
\begin{equation}
|{\cal M}(1,2,\dotsc,N)|^2 = 1+\frac{c_1^{(2)}}{Q^2}\sum_{i=1}^N p_{\perp i}^2+{\cal O}(Q^{-4})\,,
\end{equation}
where $c_1^{(2)}$ is some dimensionless, numerical coefficient.  We will discuss the potential dependence of $c_1^{(2)}$ on the number of particles $N$ later.

To construct linear relationships between terms at higher mass dimension, we can multiply the momentum conservation equations by symmetric polynomials of the minimum bias phase space coordinates.  For example, two identities that can be used to establish terms at order $1/Q^4$ are:
\begin{align}
&\sum_{i=1}^N p_{\perp i}^2\left(
\sum_{j=1}^N p_{\perp j}^2 + \sum_{j\neq k}^N p_{\perp j}p_{\perp k}\cos(\phi_j-\phi_k)
\right) = 0 \\
&
\hspace{1cm}=\sum_{i=1}^N p_{\perp i}^4 +  \sum_{j\neq k}^N p_{\perp j}^3p_{\perp k}\cos(\phi_j-\phi_k)+\sum_{i\neq j\neq k}^N p_{\perp i}^2p_{\perp j}p_{\perp k}\cos(\phi_j-\phi_k)\,,\nonumber\\
&\sum_{i=1}^N p_{\perp i}^2\left(
k^+k^--\sum_{j=1}^N p_{\perp j}^2 - \sum_{j\neq k}^N p_{\perp j}p_{\perp k}\cosh(\eta_j-\eta_k)
\right) = 0 \nonumber\\
&\hspace{1cm}=k^+k^-\sum_{i=1}^N p_{\perp i}^2-\sum_{i=1}^N p_{\perp i}^4 - \sum_{j\neq k}^N p_{\perp j}^3p_{\perp k}\cosh(\eta_j-\eta_k)-\sum_{i\neq j\neq k}^N p_{\perp i}^2p_{\perp j}p_{\perp k}\cosh(\eta_j-\eta_k)\,.\nonumber
\end{align}
Again, on the right, we have expanded the identities in terms of independent symmetric polynomials.  Then, through the first few orders, the squared matrix element can be expanded as:
\begin{align}
\label{eq:expansion0}\hspace{-0.2cm}|{\cal M}(1,2,\dotsc,N)|^2 &= 1+\frac{c_1^{(2)}}{Q^2}\sum_{i=1}^N p_{\perp i}^2+\frac{c_1^{(4)}}{Q^4}k^+k^-\sum_{i=1}^N p_{\perp i}^2+\frac{c_2^{(4)}}{Q^4}\sum_{i=1}^N p_{\perp i}^4+\frac{c_3^{(4)}}{Q^4}\sum_{i\neq j}^N p_{\perp i}^2p_{\perp j}^2\\
&
\hspace{0.5cm}+\frac{c_4^{(4)}}{Q^4}\sum_{i\neq j}^N p_{\perp i}^3p_{\perp j}\cosh(\eta_i-\eta_j)+\frac{c_5^{(4)}}{Q^4}\sum_{i\neq j}^N p_{\perp i}^2p_{\perp j}^2\cosh(2(\eta_i-\eta_j))\nonumber\\
&
\hspace{0.5cm}+\frac{c_6^{(4)}}{Q^4}\sum_{i\neq j\neq k}^N p_{\perp i}^2p_{\perp j}p_{\perp k}\cosh(\eta_i-\eta_j)\cosh(\eta_i-\eta_k)\nonumber\\
&
\hspace{0.5cm}+\frac{c_7^{(4)}}{Q^4}\sum_{i\neq j}^N p_{\perp i}^2p_{\perp j}\cos(\phi_i-\phi_j)+\frac{c_8^{(4)}}{Q^4}\sum_{i\neq j}^N p_{\perp i}^2p_{\perp j}^2\cos(2(\phi_i-\phi_j))\nonumber\\
&
\hspace{0.5cm}+\frac{c_9^{(4)}}{Q^4}\sum_{i\neq j\neq k}^N p_{\perp i}^2p_{\perp j}p_{\perp k}\cos(\phi_i-\phi_j)\cos(\phi_i-\phi_k)\nonumber\\
&
\hspace{0.5cm}+\frac{c_{10}^{(4)}}{Q^4}\sum_{i\neq  j}^N p_{\perp i}^2p_{\perp j}^2\cosh(\eta_i-\eta_j)\cos(\phi_i-\phi_j)\nonumber\\
&
\hspace{0.5cm}+\frac{c_{11}^{(4)}}{Q^4}\sum_{i\neq j\neq k}^N p_{\perp i}^2p_{\perp j}p_{\perp k}\cosh(\eta_i-\eta_j)\cos(\phi_i-\phi_k) + {\cal O}(Q^{-6})\,.
\nonumber
\end{align}
The $c_i^{(4)}$ are dimensionless numerical coefficients.  Linear relationships from conservation of momentum have already been accounted for and (hyperbolic) trigonometric identities have been used. Note also that any terms that exclusively depend on the product $k^+k^-$ can be absorbed into the function $f(k^+k^-)$ and are not included.

The numerical coefficients $c_i^{(n)}$ in general depend on the number of detected particles $N$.  For the most part, we will be agnostic as to what their precise scaling with $N$ is, but we will make the following weak, but constraining, assumption.  We assume that, in the $N\to\infty$ limit, the dependence of the coefficients $c_i^{(n)}$ on $N$ is such that the entire corresponding term in the squared matrix element remains finite.  For example, consider the term at ${\cal O}(Q^{-2})$.  With our power counting assumption that $p_\perp \sim Q/N$ and ergodicity, the kinematic dependence of this term scales like
\begin{equation}
\frac{1}{Q^2}\sum_{i=1}^N p_{\perp i}^2 \sim \frac{1}{N}\,,
\end{equation}
because there are $N$ terms in the sum and each term scales like $1/N^2$.  Thus, the coefficient $c_1^{(2)}$ can scale at worst linear in $N$ as $N\to\infty$ and the whole term will still be finite in the squared matrix element.\footnote{In matching this expansion to a short-distance description from QCD, the coefficients $c_i^{(n)}$ can generically also depend on the ratio of collision energy $Q$ to the QCD scale $\Lambda_\text{QCD}$ or pion mass $m_\pi$.  Explicitly doing the matching or resumming large logarithms of this ratio requires an effective field theory for minimum bias, which we do not construct here and leave to future work.}

Further, the assumption of ergodicity implies that in the limit that $N\to\infty$, the entire matrix element becomes a constant, independent of the event's dynamics.  Our ergodic assumption implies that the symmetric sums over terms that appear in the squared matrix element reduce to the corresponding mean values for $N\to\infty$.  For example, consider the term at ${\cal O}(Q^{-2})$ again.  The sum of the squared transverse momentum of all particles in the event returns the mean square transverse momentum $\langle p_\perp^2\rangle$ times $N$, as $N\to\infty$.  Because the sum consists of $N$ terms, the variance of the sum is also of order $N$ and so the whole expression takes the form
\begin{equation}
\lim_{N\to\infty}\sum_{i=1}^N p_{\perp i}^2 \to N\langle p_\perp^2\rangle + {\cal O}\left(\sqrt{N}\langle p_\perp^2\rangle\right)\,.
\end{equation}
In the strict $N\to\infty$ limit, the variance of the sum vanishes, reducing to a constant value on all of phase space.  Every term in the squared matrix element reduces to its mean as $N\to\infty$ by ergodicity, and so the whole squared matrix element itself becomes a constant on the $N$-body phase space as $N\to\infty$.  We will exploit this limit of the matrix element in our predictions of the following section.

While constructive, this $1/Q$ expansion isn't exactly the expansion that is natural with our power counting.  Terms at a fixed order in $1/Q$ do not in general have a homogeneous scaling with $1/N$.  For example, at ${\cal O}(Q^{-4})$, the term
\begin{equation}
\frac{1}{Q^4}k^+k^-\sum_{i=1}^N p_{\perp i}^2 \sim \frac{1}{N}\,,
\end{equation}
while the term
\begin{equation}
\frac{1}{Q^4}\sum_{i=1}^N p_{\perp i}^4 \sim \frac{1}{N^3}\,,
\end{equation}
as $N\to\infty$.  The particular form of higher-order terms in $1/N$ is subtle to determine because, in general, terms that arise in the $1/Q$ expansion don't have a homogeneous scaling in $1/N$.  For example, assuming that $k^+k^-$ is homogeneous in powers of $N$ with $k^+k^-\sim N^0$, then from \Eq{eq:momconserv} the term with hyperbolic cosine is not homogeneous in $N$:
\begin{equation}
\sum_{i\neq j}^N p_{\perp i}p_{\perp j}\cosh(\eta_i-\eta_j)\sim {\cal O}(N^0) + {\cal O}(N^{-1})\,.
\end{equation}
This follows because of the assumption that $\eta_i\sim 1$ and so $\cosh(\eta_i-\eta_j)\sim 1$, the transverse momentum $p_{\perp i}\sim 1/N$ and that the sum consists of $N^2-N$ terms.  We do not attempt to determine the general form of the $1/N$ expansion of the squared matrix element here, although a systematic approach could be pursued along the lines of \Refs{Henning:2015daa,Henning:2015alf,Henning:2017fpj,Henning:2019enq,Henning:2019mcv,Graf:2020yxt,Melia:2020pzd}.  However, what we will do in \Sec{sec:datacomp}, when using the power counting and symmetries to understand collider data, is identify the $1/N$ expansion for particular subsets of terms in the squared matrix element that are relevant for the specific measurements of interest that we consider in this work.

One important point to make at this stage is that we are considering fixed, but large, number of particles $N$.  That is, our expansion can make predictions for observables that survive in the $N\to\infty$ limit or for observables that are conditioned on $N$.  We will see a number of examples of such observables in \Sec{sec:datacomp}.  However, with our present formulation of the cross section for minimum bias, we cannot make predictions for which the relative probability of different numbers of detected particles is important.  Perhaps the simplest such observable is just the multiplicity $N$ of detected particles itself.  We leave an understanding of particle multiplicity within this framework to future work.

\subsection{Where are Jets?}

Jets, while the most ubiquitous phenomena of high-energy QCD, are not described within the framework of our minimum bias expansion of the cross section.  From the perspective of our goals with this expansion, this is a good thing, as it implies that the expansion focuses on the physics of interest, namely the low-energy dynamics of QCD.  Specifically, the presence of high-energy jets in the detected final state particles violates the power counting assumption that the mean particle transverse momentum, $\langle p_\perp\rangle$, is representative of all particles' transverse momentum.  Jets are collimated streams of particles which carry an order-1 fraction of the jet's energy.  Very low energy particles fill in the regions at wide angles from the jets.  The distribution of particle transverse momentum then has two dominant populations corresponding to high energy or very low energy particles.  Of course, for any distribution of $N$ particles with total energy $Q$, the mean transverse momentum $\langle p_\perp\rangle \sim Q/N$.  However, when jets are present in the final state, the root mean square transverse momentum $\sqrt{\langle p_\perp^2\rangle}$ will be dominated by the high energy particles of the jets, and so we find the hierarchy that $\sqrt{\langle p_\perp^2\rangle}\gg \langle p_\perp\rangle \sim Q/N$.  This then means that jets are not described by any low-order truncation of either the $1/Q$ or $1/N$ expansion of the squared matrix element. 

Of course, this is not surprising as jets arise because of the approximate scale invariance of QCD at high energies, in addition to the small coupling $\alpha_s$.  Even approximate scale invariance is not a symmetry of our definition of minimum bias, and so is not exploited in the expansion of the squared matrix element.  Further, the smallness of $\alpha_s$ for controlling jet dynamics means that jets can be described by a perturbative effective field theory, known as soft-collinear effective theory \cite{Bauer:2000yr,Bauer:2001ct,Bauer:2001yt,Bauer:2002nz}.  For minimum bias, because the only relevant energy scale is the mean transverse momentum $\langle p_\perp\rangle \sim Q/N\ll Q$, we cannot assume that $\alpha_s$ is small so as to formulate a perturbative effective field theory.  

Nevertheless, even in the presence of collimated jets, the low energy particles that fill the detector away from the jets should exhibit many of the properties we have already discussed.  Namely, as long as the number $N$ of low energy particles not in the jets is large, we still expect their net transverse momentum to be approximately 0, up to the relative $1/\sqrt{N}$ standard deviation of a random walk.  However, jets in the final state introduce new, relevant directions that spontaneously break the symmetries of minimum bias as well as introduce correlations that would manifest themselves in the form of the squared matrix element.  As discussed, such correlations would not be captured to low orders in our expansion of the cross section. 

%%%%%%%%%%%%%%%%%%%%%%%%%%%%%%%%%%
\section{Interpretation of Data}\label{sec:datacomp}
%%%%%%%%%%%%%%%%%%%%%%%%%%%%%%%%%%

In this section, we exploit this expansion of the cross section for minimum bias events to understand and re-interpret collider data from this perspective.  Because the expansion as we have developed it thus far can describe any identical hadron or nucleus scattering, we apply it to understand data from both pp and heavy ion collisions.  Particularly in the case of heavy ion collisions, experimental data are very often interpreted or expressed in terms of strictly unobservable quantities as established in some model of the collision, like the centrality or the number of participating nucleons.  By contrast, our expansion is expressed exclusively in terms of directly observable quantities, like the total number of detected particles or the total observed final state energy.

%%%%%%%%%%%%%%%%%%%%%%%%%%%%%%%%%%
\subsection{Pseudorapidity Distributions}
%%%%%%%%%%%%%%%%%%%%%%%%%%%%%%%%%%

The first observable that we consider is the pseudorapidity distribution of observed final state particles.  This can be calculated directly from the form of the minimum bias cross section we established in \Eq{eq:xsecmaster}.  From the permutation symmetry of the particles, every particle has the same pseudorapidity distribution $p(\eta)$, so we can just fix the particle of interest to be particle 1.  Then, we have
\begin{align}
p(\eta)&\sim \int_0^Q dk^+ \int_0^Q dk^- \int \prod_{i=1}^N\left[p_{\perp i}\, dp_{\perp i}\, d\eta_i\, \frac{d\phi_i}{2\pi}  \right]\, f\left(
k^+k^-
\right)\, |{\cal M}(1,2,\dotsc,N)|^2 \\
&
\hspace{1cm}\times \delta(\eta-\eta_1)\, \delta\left(
k^- - \sum_{i=1}^N p_{\perp i}e^{\eta_i}
\right)\, \delta\left(
k^+ - \sum_{i=1}^N p_{\perp i}e^{-\eta_i}
\right)\,\delta^{(2)}\left(
\sum_{i=1}^N \vec p_{\perp i}
\right)\nonumber\,.
\end{align}
To proceed, we will make a number of simplifying assumptions.  First, we will work in the $N\to\infty$ limit in which the squared matrix element $|{\cal M}(1,2,\dotsc,N)|^2$ reduces to a constant on phase space, as discussed earlier.  Because this constant is just an overall scaling, we can take it to be 1, for simplicity.  With this assumption, we can then determine the pseudorapidity distribution on flat phase space $p_\text{flat}(\eta)$, in the large-$N$ limit.  We have
\begin{align}\label{eq:flatmaster}
p_\text{flat}(\eta) &\sim \lim_{N\to\infty}\int \prod_{i=1}^N \left[
p_{\perp i}\, dp_{\perp i}\, d\eta_i\, \frac{d\phi_i}{2\pi}
\right]\, \delta(\eta - \eta_1)\\
&
\hspace{2cm}
\times \delta\left(
k^--\sum_i p_{i\perp}e^{\eta_i}
\right)\delta\left(
k^+-\sum_i p_{i\perp}e^{-\eta_i}
\right)\, \delta^{(2)}\left(
\sum_i \vec p_{i\perp}
\right)\nonumber\\
&\propto \lim_{N\to\infty}\int dp_{\perp 1}\, p_{\perp 1}\, d\eta_1\, \delta(\eta - \eta_1)\left[
(k^+-p_{\perp 1}e^{-\eta_1})(k^--p_{\perp 1}e^{\eta_1})-p^2_{\perp 1}
\right]^N\nonumber\\
&\propto\lim_{N\to\infty}\int  dp_\perp\,p_\perp \left(
1-\frac{k^+e^{\eta}+k^-e^{-\eta}}{k^+k^-}\,p_\perp
\right)^N\nonumber\\
&=\int_0^\infty  dp_\perp\,p_\perp \, e^{-\frac{k^+e^{\eta}+k^-e^{-\eta}}{k^+k^-}\,Np_\perp}\nonumber\\
&=\frac{(k^+k^-)^2}{N^2}\left(
k^+e^{\eta}+k^-e^{-\eta}
\right)^{-2}\,.\nonumber
\end{align}
To get this expression, we used the large-$N$ limit of the volume of phase space from \Eq{eq:psvol} and ignore overall factors.  The normalized probability distribution of particle pseudorapidity on flat phase space is then
\begin{equation}\label{eq:flateta}
p_\text{flat}(\eta) = 2k^+k^-\left(
k^+e^{\eta}+k^-e^{-\eta}
\right)^{-2}\,.
\end{equation}
In the large-$N$ limit, this can also be directly derived from assuming that all particles have a distribution flat in $\cos\theta$, the polar angle from the beam.  Note that this large-$N$ limit ignores momentum conservation in transforming to the exponential integrand.

We can then use this result to determine the observed pseudorapidity distribution $p(\eta)$ smeared against the function $f(k^+k^-)$.  We have
\begin{align}\label{eq:etadistsmear}
p(\eta)&=\frac{1}{Q^2}\int_0^Q dk^+\int_0^Q dk^-\, f\left(k^+k^-\right)
\, p_\text{flat}(\eta) \\
&= \frac{1}{Q^2}\int_0^Q dk^+\int_0^Q dk^-\, f\left(k^+k^-\right)
\, 2k^+k^-\left(
k^+e^{\eta}+k^-e^{-\eta}
\right)^{-2}\nonumber\\
&=\int_0^1 dx\, f(x)\, \frac{1-x^2}{1+x^2+2x \cosh(2\eta)}
\,.\nonumber
\end{align}
In the final line, we have set $xQ^2=k^+k^-$ and integrated over the pseudorapidity of the system of final state particles in the lab frame.  Written in this way, $f(x)$ is itself a probability distribution whose normalization is inherited from its expression in terms of $k^+$ and $k^-$:
\begin{align}
1&=\frac{1}{Q^2}\int_0^Q dk^+ \int_0^Q dk^-\, f(k^+k^-) = \int_0^1 dx\, \log\frac{1}{x}\, f(x)\,.
\end{align}
We also note that because $f(x)\geq 0$ on $x\in[0,1]$, the pseudorapidity distribution $p(\eta)$ monotonically decreases as $|\eta|$ increases away from $\eta=0$.  The integral in the final line of \Eq{eq:etadistsmear} is also dominated by the region for $0\leq x\lesssim 1/\cosh(2\eta)$.

To go further, we need an explicit form for the function $f(x)$.  As we noted earlier, $f(x)$ should be analytic on $x\in[0,1]$ because it is a physical squared matrix element.  Also, because of the collinear singularity of QCD, we expect a very flat distribution of pseudorapidity over a wide range.  From our expression for the pseudorapidity distribution $p(\eta)$ in \Eq{eq:etadistsmear}, we note that if $f(x) = \delta(x)$, then $p(\eta)$ is flat, but not normalizable.  So, combining these observations suggests that $f(x)$ should be highly-peaked at $x = 0$ and analytic.  This motivates the following form:
\begin{align}\label{eq:cutffunc}
f(k^+k^-) = \frac{n+1}{H_{n+1}}\left(1-\frac{k^+k^-}{Q^2}\right)^n\simeq \frac{n}{\gamma_E+\log\, n}\,e^{-n\frac{k^+k^-}{Q^2}}\,,
\end{align}
where $n\gg 1$ ensures strong peaking at $x=0$, $H_n$ is the harmonic number and $\gamma_E$ is the Euler-Mascheroni constant.\footnote{We have currently described the function $f(k^+k^-)$ as a function that corresponds to smearing the flat phase space distribution.  However, it can equivalently be interpreted as a squared matrix element $|{\cal M}|^2$ for the $N$ detected particles using the relationship from \Eq{eq:momconserv}.  With this interpretation, the squared matrix element is
\begin{equation}
|{\cal M}|^2 \sim e^{-n\frac{k^+k^-}{Q^2}} \sim \exp\left[
-n\sum_{i\neq j}^N \frac{p_{\perp i}p_{\perp j}}{Q^2}\cosh(\eta_i-\eta_j)
\right]\,,
\end{equation}
to leading order in $1/N$ in the exponent.  Of course, regardless of the interpretation, we find the same predictions for the desired measured quantities.} We discuss the independence of our results on the precise form of this function shortly.

To determine the value $n$, we must use data.  Pseudorapidity distributions have been measured extensively in the ATLAS \cite{ATLAS:2011ag,Aad:2012mfa,Aad:2015wga,Aad:2016mok}, CMS \cite{Khachatryan:2010xs,Khachatryan:2010us,Chatrchyan:2011av,Chatrchyan:2014qka,Khachatryan:2015jna,Chatrchyan:2011pb,Sirunyan:2019cgy} and ALICE \cite{Aamodt:2009aa,Abbas:2013bpa,Adam:2015kda,Adam:2015pza,Adam:2016ddh,Acharya:2018hhy,Aamodt:2010my,Abelev:2013ala,ALICE:2015qqj,Acharya:2018qsh,Acharya:2018eaq} experiments at the LHC, in both pp and heavy ion collisions.  From those data, we can determine the parameter $n$ by noting the following.  The mean value of $x$ from \Eq{eq:cutffunc} is roughly $1/n$, which is the region that dominates the integral.  We call $\eta_{1/2}$ where the pseudorapidity distribution is approximately half of its value at $\eta = 0$ which occurs when $2x \cosh(2\eta_{1/2}) \sim 1$; or when
\begin{equation}\label{eq:etamaxn}
n = 2\cosh(2\eta_{1/2})\,.
\end{equation}
From the 8 TeV LHC pp experiment data that follows, for instance, the value of $\eta_{1/2}\sim 6$, and so $n \sim 1.6\times 10^5$. The parameter $n$ has implicit $Q$ dependence, at least by the connection to maximum value of $\eta$.

We emphasize that in what follows that the fit above to the tail around $|\eta|\gtrsim \eta_{1/2}$ is not important for the observables that are captured in our approach; that is, any step-like function with a cutoff around $\eta_{1/2}$ would suffice.  Kinematic observables binned in $N$ and  at fixed $Q$ in the region of validity of this effective framework are insensitive as $n\to\infty$; we will see this explicitly below with the single particle $p_\perp$ distribution, where we also give a quantitative estimate for the validity condition. In order to correctly capture scalings with $Q$, the value of $n$ is important in that it captures the value of $\eta_{1/2}$, which scales with $Q$. Any other step-like function, however, would capture the same scaling and, at least at large enough $Q$, this scaling should be independent of the precise form of the tail.

\begin{figure}
\begin{center}
\includegraphics[width=9cm]{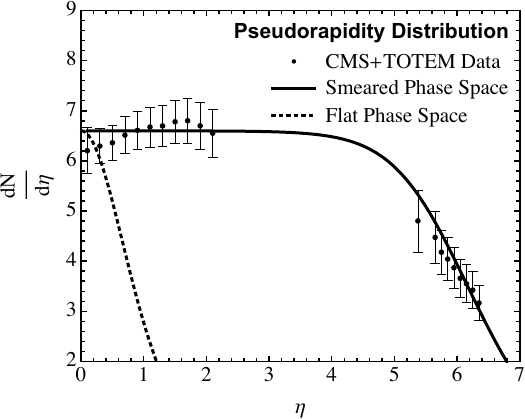}
\caption{Comparison of particle density in pseudorapidity between 8 TeV pp collider data from the CMS and TOTEM experiments \cite{Chatrchyan:2014qka} to our prediction of \Eq{eq:etadistsmear} (solid) and the flat phase space distribution of \Eq{eq:flateta} (dashed).  The data only include charged particles with transverse momentum greater than 40 MeV and our prediction uses the form of smearing function $f(x)$ in \Eq{eq:cutffunc} with $n = 1.6\times 10^5$. 
\label{fig:etacomp}
}
\end{center}
\end{figure}

The above established value of $n$ can then be used to compare our predicted pseudorapidity distribution to data.  This is done in \Fig{fig:etacomp} where we compare our prediction to data from the CMS and TOTEM experiments \cite{Chatrchyan:2014qka} of the charged particle number density as a function of pseudorapidity, $dN/d\eta$, in 8 TeV pp collisions.  Our prediction for the pseudorapidity distribution is just a normalized probability distribution, so we multiply by a factor to fit data.  As mentioned earlier, our present formulation of the expansion of the minimum bias cross section does not enable us to predict the multiplicity distribution, so we can't predict the overall normalization here.  In general, we find good agreement between our prediction and data, with a noticeable lack of a dip in our prediction near $\eta = 0$.  Hadrons are of course massive particles and there is a distinction between their rapidity and pseudorapidity which is manifest as the dip in the pseudorapidity distribution.  We assumed that the transverse momentum of particles is much larger than their mass, so we ignore the pseudo/rapidity distinction.  However, the data include particles with transverse momentum above 40 MeV which includes hadrons with transverse momentum comparable to their mass.

%%%%%%%%%%%%%%%%%%%%%%%%%%%%%%%%%%
\subsection{Transverse Momentum Distributions}
%%%%%%%%%%%%%%%%%%%%%%%%%%%%%%%%%%

We now turn to understanding transverse momentum distributions in pp collisions.  The set-up for this analysis will be the same as that for pseudorapidity.  We take the squared matrix element for the detected particles $|{\cal M}|^2 = 1$ and take the large-$N$ limit of phase space.  The first steps are therefore very similar to that for pseudorapidity, so we won't repeat them here.  From \Eq{eq:flatmaster}, the flat phase space distribution of the transverse momentum is
\begin{equation}
p_\text{flat}(p_\perp) \propto p_\perp \int_{-\infty}^\infty d\eta\, e^{-\frac{k^+e^{\eta}+k^-e^{-\eta}}{k^+k^-}\,Np_\perp}= p_\perp \, K_0\left(
\frac{2N p_\perp}{\sqrt{k^+k^-}}
\right)\,,
\end{equation}
where $K_0(z)$ is a modified Bessel function.  The unit normalized distribution is
\begin{equation}\label{eq:flatpspt}
p_\text{flat}(p_\perp) = \frac{4N^2 p_\perp}{k^+k^-}\, K_0\left(
\frac{2Np_\perp}{\sqrt{k^+k^-}}
\right)\,.
\end{equation}
With this result, the distribution smeared with the function $f(k^+k^-)$ is, in general,
\begin{align}
p(p_\perp)&=\frac{1}{Q^2} \int_0^Q dk^+\int_0^Q dk^-\, f\left(k^+k^-\right)\, p_\text{flat}(p_\perp)\\
&=\frac{1}{Q^2} \int_0^Q dk^+\int_0^Q dk^-\, f\left(k^+k^-\right)\, \frac{4N^2 p_\perp}{k^+k^-}\, K_0\left(
\frac{2Np_\perp}{\sqrt{k^+k^-}}
\right)\nonumber\\
&=
\frac{4N^2 p_\perp}{Q^2}\int_0^1 dx\, \log\frac{1}{x}\, f(x)\, \frac{1}{x}\, K_0\left(
\frac{2Np_\perp}{Q\sqrt{x}}
\right)
\nonumber\,.
\end{align}

Unlike the pseudorapidity distribution, the transverse momentum distribution depends explicitly on the number of detected particles $N$.  This number fluctuates event-by-event and we do not predict the multiplicity distribution. So, instead, we will re-write this distribution in terms of the mean transverse momentum, which we can calculate and is unique over an ensemble of collision events.  This mean is
\begin{align}
\langle p_\perp \rangle &= \int_0^\infty dp_\perp\, p_\perp \, p(p_\perp) =\frac{4N^2}{Q^2}\int_0^1 dx\, \log\frac{1}{x}\, f(x)\, \frac{1}{x} \int_0^\infty dp_\perp \,p_\perp^2\,  K_0\left(
\frac{2Np_\perp}{Q\sqrt{x}}
\right)\\
&=\frac{\pi}{4}\frac{Q}{N}\int_0^1 dx\, \log\frac{1}{x}\,f(x)\, \sqrt{x} \equiv \frac{\pi}{4}\frac{Q}{N} \langle \sqrt{x}\rangle
\nonumber\,.
\end{align}
Substituting this expression for $N$, the transverse momentum distribution is then
\begin{align}
p(p_\perp)&=\frac{\pi^2 \langle \sqrt{x}\rangle^2}{4\langle p_\perp\rangle^2 }\,p_\perp\int_0^1 dx\, \log\frac{1}{x}\, f(x)\, \frac{1}{x}\, K_0\left(
\frac{\pi \langle \sqrt{x}\rangle}{2\langle p_\perp\rangle\sqrt{x}}\, p_\perp
\right)
\nonumber\,.
\end{align}
With $f(x)$ from \Eq{eq:cutffunc}, the mean transverse momentum is
\begin{equation}\label{eq:meanpt}
\langle p_\perp\rangle = \frac{\pi}{4}\frac{Q}{N} \langle \sqrt{x}\rangle \simeq \frac{\pi^{3/2}Q}{8\sqrt{n}N}\,,
\end{equation}
in the $n\to\infty$ limit. The $p_\perp$ distribution with this form of $f(x)$ is
\begin{align}\label{eq:perpdistfit}
p(p_\perp)&=
\frac{\pi^3 p_\perp}{16 \langle p_\perp\rangle^2}\frac{1}{\gamma_E+\log\,n }\int_0^1 dx\, \log\frac{1}{x}\, e^{-nx}\, \frac{1}{x}\, K_0\left(
\frac{\pi^{3/2}p_\perp}{4\langle p_\perp\rangle\sqrt{x n}}
\right)\,.
\end{align}
Data of the transverse momentum distribution are often displayed as a number density per phase space volume, and so what we will plot is actually
\begin{equation}
\frac{1}{2\pi p_\perp}\frac{dN}{dp_\perp} \propto \frac{p(p_\perp)}{p_\perp}\,.
\end{equation}
Note that once we determine $f(x)$ from pseudorapidity data, the prediction of the transverse momentum distribution only depends on a single parameter, $\langle p_\perp\rangle$.  Additionally, a cut on the maximum pseudorapidity of particles that contribute to this distribution can be incorporated, as such a cut is always imposed in data.  However, any such cut that is relevant experimentally has an exceedingly small effect on the transverse momentum distribution, so we will not include it in what follows.

The reason for effective independence on a pseudorapidity cut is as follows.  In the large $n$ limit, the transverse momentum distribution of \Eq{eq:perpdistfit} is itself independent of $n$.  The Bessel function has an asymptotic form of
\begin{equation}
K_0(z) \to \sqrt{\frac{\pi}{2z}}\,e^{-z}\,,
\end{equation}
for $z\gg1$.  With this approximation, the distribution is
\begin{align}
p(p_\perp)&\sim
\sqrt{p_\perp} \int_0^1 dx\, \log\frac{1}{x}\, \frac{e^{-nx-\frac{\pi^{3/2}p_\perp}{4\langle p_\perp\rangle\sqrt{x n}}}}{x^{3/4}}\,,
\end{align}
ignoring overall constant factors.  Now, with $n\gg1$, we can saddle-point approximate the exponential.  The value of $x$ for which the exponent factor is minimized is
\begin{equation}
x_{\min} = \frac{\pi}{4n}\frac{p_\perp^{2/3}}{\langle p_\perp\rangle^{2/3}}\,.
\end{equation}
Just setting $x$ in the integrand equal to this minimum value, taking $n\to\infty$ and ignoring non-exponential factors, we have
\begin{align}\label{eq:asymppt}
p(p_\perp)&\sim e^{-\frac{3\pi}{4}\frac{p_\perp^{2/3}}{\langle p_\perp\rangle^{2/3}}}\,.
\end{align}
As the value of $n$ in turn determines the maximum value of pseudorapidity according to \Eq{eq:etamaxn}, as long as $n$ is large enough, any dependence on a pseudorapidity cut is eliminated.

This probability distribution is like the Boltzmann factor for the ``gas'' of $N$ detected particles.  Hence, they have a dispersion relation of $\omega \propto k_\perp^{2/3}$.  Fractional dispersion relations are very strange, but can arise from integrating out a gapless subsystem of a larger system \cite{Watanabe:2014zza}.  This is essentially what we did here: the beam remnants were gapless, but we exclusively wanted an effective description of the detected particles.  Integrating out gapless modes introduces non-locality in the effective theory, which is manifested as a non-analytic dispersion relation.  In particular, note that the non-analyticity is not present in the flat phase space distribution of \Eq{eq:flatpspt}.  The asymptotic form of the Bessel function implies that
\begin{equation}
p_\text{flat}(p_\perp)\sim e^{-\frac{\pi}{2}\frac{p_\perp}{\langle p_\perp\rangle}}\,,
\end{equation}
ignoring overall non-exponential factors.  No gapless modes are integrated out on flat phase space, hence the dispersion relation $\omega \propto k_\perp$ is analytic.  The particular form of the dispersion relation implied by \Eq{eq:asymppt} may provide {a hint towards the construction of an effective field theory for minimum bias based on spontaneously broken symmetry breaking patterns, but we leave that to future work.}

We can also verify the consistency of our transverse momentum prediction, within the framework of our assumed power counting.  The second moment of the transverse momentum is
\begin{align}
\langle p_\perp^2 \rangle &= \int_0^\infty dp_\perp\, p_\perp^2 \, p(p_\perp) =\frac{4N^2}{Q^2}\int_0^1 dx\, \log\frac{1}{x}\, f(x)\, \frac{1}{x} \int_0^\infty dp_\perp \,p_\perp^3\,  K_0\left(
\frac{2Np_\perp}{Q\sqrt{x}}
\right)\\
&=\frac{Q^2}{N^2}\int_0^1 dx\, \log\frac{1}{x}\,f(x)\,x
= \frac{Q^2}{n N^2}\,,\nonumber
\end{align}
where the final equality is the large $n$ limit of the second moment of the distribution of \Eq{eq:perpdistfit}.  Then, the root mean square and expectation value of the transverse momentum are related by
\begin{equation}
\sqrt{\langle p_\perp^2 \rangle} = \frac{8}{\pi^{3/2}}\langle p_\perp \rangle \simeq 1.44 \langle p_\perp \rangle\,,
\end{equation}
satisfying our power counting of $\sqrt{\langle p_\perp^2 \rangle}\sim\langle p_\perp \rangle$ and demonstrating consistency of our prediction.

\begin{figure}[t]
\begin{center}
\includegraphics[width=9cm]{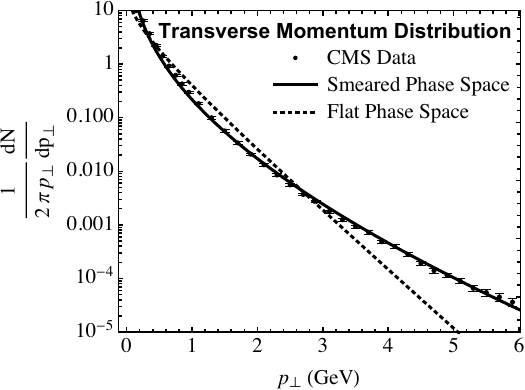}
\caption{Transverse momentum distribution of charged particles in minimum bias events in $\sqrt{s}=7$ TeV pp collisions at CMS from \Ref{Khachatryan:2010us} compared to our prediction.  The data have an $|\eta|< 2.4$ cut on the charged particles.  Our prediction from the asymptotic expression of \Eq{eq:asymppt} (solid) uses $\langle p_\perp\rangle = 0.65$ GeV, and also plotted is the distribution on flat phase space of \Eq{eq:flatpspt} (dashed).
\label{fig:ptpred}}
\end{center}
\end{figure}

With this prediction for the transverse momentum distribution, we can compare it to collider data.  \Fig{fig:ptpred} shows the transverse momentum distribution of charged particles in minimum bias events in $\sqrt{s}=7$ TeV pp collisions at CMS from \Ref{Khachatryan:2010us}.  A pseudorapidity cut of $|\eta|<2.4$ is also imposed, but, as mentioned earlier, that will not matter for our prediction.  This figure also plots the asymptotic form of our prediction from \Eq{eq:asymppt} where we have set the parameter $\langle p_\perp\rangle = 0.65$ GeV, as well as the transverse momentum distribution on flat $N$-body phase space.  Our prediction is in remarkably good agreement with the data over the plotted range of transverse momentum.  We want to emphasize that once the form of the smearing function $f(x)$ is determined from pseudorapidity data, our prediction for the transverse momentum distribution depends on only one parameter, the mean value of the transverse momentum.

While we are unable to directly predict the multiplicity distribution from this framework, we can nevertheless demonstrate consistency between parameters in the expansion of the cross section with measured values of the mean multiplicity.  From \Eq{eq:meanpt}, the expected value of $N$ with our choice for the form of the function $f(x)$ is
\begin{equation}\label{eq:expmult}
N  \simeq \frac{\pi^{3/2}Q}{8\sqrt{n}\langle p_\perp\rangle}\,.
\end{equation}
Fitting the 7 TeV data from CMS in \Fig{fig:ptpred} fixes $\langle p_\perp\rangle = 0.65$ GeV, and from the 8 TeV CMS+TOTEM data of the pseudorapidity distribution from \Fig{fig:etacomp}, we fit $n=1.6\times 10^5$.  While these data are from different collision energies and so are perhaps not directly comparable and interpretable from one to the other, we only anticipate logarithmic dependence on center-of-mass collision energy, so their distinction should be minimal when applied to our (rather coarse) scaling predictions.  Taking these values along with $Q = 8$ TeV, the expected value of the detected multiplicity from \Eq{eq:expmult} is then
\begin{equation}
N \simeq 21\,.
\end{equation}
While this is a very simple and crude prediction, it is nevertheless in the same order-of-magnitude as the number of observed charged particles from \Fig{fig:etacomp}, for example.  In that figure, the number density of charged particles with transverse momentum greater than 40 MeV is roughly 6.5 per unit pseudorapidity for $|\eta| \lesssim 5.5$.  So, there are roughly 72 charged particles in each event.  Our prediction is about a factor of 3 smaller, which is likely accounted for by the transverse momentum cut.  40 MeV is less than the mass of the pion, and so violates an assumption of our power counting.  Increasing this cut would correspondingly decrease the number of detected particles, while not affecting our fit value for $\langle p_\perp\rangle$.

Further, the expression for multiplicity $N$ from \Eq{eq:expmult} implies a non-trivial dependence on the center-of-mass collision energy.  First, note that the value of $n$ in the form of the function $f(x)$ is related to the collision energy $Q$ through \Eq{eq:etamaxn}:
\begin{equation}
\eta_{\max}\simeq \log\frac{Q}{p_{\perp\text{cut}}} \simeq \log \,n\,.
\end{equation}
Here, $p_{\perp\text{cut}}$ is the experimental lower bound on detected particle transverse momentum.  Then, as long as the dependence of the mean transverse momentum $\langle p_\perp\rangle$ on the collision energy $Q$ is relatively weak, the multiplicity scales with a fractional power of $Q$:
\begin{equation}
N \sim Q^{1/2} = s^{1/4}\,.
\end{equation}
This particular fractional power scaling is a prediction of the Fermi-Landau model \cite{Fermi:1950jd,Landau:1953gs,Belenkij:1955pgn,Wong:2008ex}.  In the context of our analysis here, we note that it is a consequence of our particular choice of the function $f(x)$ in the smeared cross section.  Additionally, the inclusion of a squared matrix element with non-trivial dependence on the detected particles will affect this scaling.  Data prefer a slightly smaller power-law scaling of the multiplicity, e.g., \Ref{ALICE:2015qqj} in which a power law of $s^{0.11}$ fits the charged particle multiplicity over decades of collision energies.  Nevertheless, this simple result within the context of our large-$N$ expansion suggests that an appropriate squared matrix element could fit the data.  We leave a more detailed analysis of the collision energy dependence of the multiplicity to future work.

Data of transverse momentum distributions in minimum bias are often compared to Tsallis distributions \cite{Tsallis:1987eu,Wilk:2008ue,Biro:2008hz} that assume there are fluctuations in the $N$ particle final state that are quantified by a non-extensive form of entropy.  This is an intriguing interpretation and the success of such models may point to fundamental fractal-like structure of particles produced in minimum bias collisions.  While not inconsistent with this interpretation, our expansion of the minimum bias cross section has a more mundane understanding as a consequence of the symmetries of these collision events.  Further, the Tsallis distribution reduces to a power law at large transverse momentum, while our smeared prediction is exponential, though dependent on a fractional power of transverse momentum.  A detailed study to distinguish the consequences of these two (or other) models of minimum bias dynamics may reveal the microscopic description of these events and lead to an effective field theory in which precision calculations can be performed.

\subsubsection{Limit of Large-$N$ Expansion}

\begin{figure}[t]
\begin{center}
\includegraphics[width=9cm]{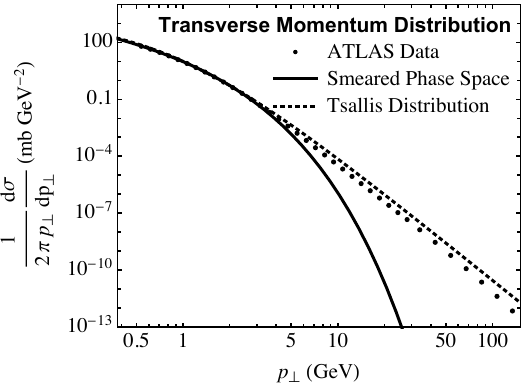}
\caption{
Transverse momentum distribution of charged particles in minimum bias events in $\sqrt{s}=2.76$ TeV pp collisions at ATLAS from \Ref{Aad:2015wga} compared to our asymptotic prediction and a Tsallis distribution.  The data have an $|\eta|< 2$ cut on the charged particles.  In our prediction of \Eq{eq:asymppt}, we set the parameter $\langle p_\perp\rangle = 0.5$ GeV and the parameters of the Tsallis distribution are $n=6.6$ and $T = 0.12$ GeV.
\label{fig:highptpred}}
\end{center}
\end{figure}

With data that extends to higher values of transverse momentum, we can see the limit of our prediction, with the assumptions we have made thus far.  In particular, we have assumed that the squared matrix element is just 1, which is its expression at lowest order in the $1/N$ expansion.  However, at higher transverse momentum, higher order terms in the squared matrix element become more important and may be necessary to describe the distribution.  In Fig.~\ref{fig:highptpred}, we compare charged particle transverse momentum distribution data from $\sqrt{s}=2.76$ TeV pp collisions to our asymptotic prediction of \Eq{eq:asymppt} and a Tsallis distribution.  This Tsallis distribution takes the form
\begin{equation}
\frac{1}{2\pi p_\perp}\frac{d\sigma}{dp_\perp}\propto \left(
1+\frac{p_\perp}{nT}
\right)^{-n}\,,
\end{equation}
for parameters $n$ and $T$. In the plot, we have set the parameters in our prediction and the Tsallis distribution to be $\langle p_\perp\rangle = 0.5$ GeV, $n=6.6$ and $T = 0.12$ GeV.  At low transverse momentum, both our prediction and the Tsallis distribution follow the data extremely well, but around 5 GeV, our prediction exponentially drops, while the Tsallis distribution largely follows the power-law distribution at high transverse momentum.  It would be interesting to study the effect of maintaining momentum conservation in our prediction and non-trivial terms in the matrix element to see if they can reproduce the high-$p_\perp$ tail, but we leave that to future work.  

However, with the approximations that we have made thus far, we can estimate where our description of the transverse momentum distribution should break down.  In the derivation of our results, an important approximation that we made in the large-$N$ limit was that momentum conservation could be ignored, which in turn resulted in an exponential appearing in the calculation of distributions on flat phase space in \Eq{eq:flatmaster}.  By carefully analyzing this approximation, we can identify where it  breaks down.  First, the large-$N$ flat phase space factor can be written as
\begin{align}
\left(
1-\frac{k^+e^\eta  + k^-e^{-\eta}}{k^+k^-}p_\perp
\right)^N &= \exp\left[
N\,\log\left(
1-\frac{k^+e^\eta  + k^-e^{-\eta}}{k^+k^-}p_\perp
\right)
\right]\\
&=\exp\left[
N\left(
-\frac{k^+e^\eta  + k^-e^{-\eta}}{k^+k^-}p_\perp-\frac{1}{2}\frac{(k^+e^\eta  + k^-e^{-\eta})^2}{(k^+k^-)^2}p^2_\perp + \cdots
\right)
\right]
\nonumber\,.
\end{align}
So far, this is exact, but higher-order terms in the exponent can be safely ignored only if
\begin{equation}
p_\perp \ll \frac{k^+k^-}{k^+e^\eta  + k^-e^{-\eta}} \,.
\end{equation}
Recall that $k^+k^-$ is the total squared energy of the detected particles.  In coordinates that we used earlier, we expressed
\begin{align}
&x = \frac{k^+k^-}{Q^2}\,, &\Delta\eta =\frac{1}{2}\, \log\frac{k^-}{k^+}\,,
\end{align}
and so the limit on transverse momentum is
\begin{equation}
p_\perp \ll \frac{k^+k^-}{k^+e^\eta  + k^-e^{-\eta}} = \frac{\sqrt{x}\,Q}{2\cosh(\eta - \Delta\eta)}\,,
\end{equation}
where $x\in[0,1]$ and 
\begin{equation}
\frac{1}{2}\,\log\, x \leq \Delta\eta \leq \frac{1}{2}\,\log\frac{1}{x}\,.
\end{equation}

To set an upper bound for the transverse momentum, we first maximize over $x$ and $\Delta\eta$, for which $x = 1$ and $\Delta \eta = 0$:
\begin{equation}
p_\perp \ll \frac{Q}{\cosh \eta}\,.
\end{equation}
That is, the energy of an individual particle must be parametrically less than the collision energy.  We had found that for pp collisions with energy of several TeV, the maximum pseudorapidity was about $\eta_{\max}\sim 6$.  For $Q = 2.76$ TeV from \Fig{fig:highptpred}, the transverse momentum then must be smaller than
\begin{align}
p_\perp \ll \frac{Q}{\cosh \eta_{\max}} \sim 14\text{ GeV}\,.
\end{align}
This estimate of the limit agrees well with \Fig{fig:highptpred}, in which our prediction diverges from data well before 14 GeV.  Conversely, for a fixed transverse momentum $p_\perp$, the maximum pseudorapidity is logarithmically related to the collision energy $Q$, which is well known \cite{Feynman:1969ej,Wilson:1970zzb}.

%%%%%%%%%%%%%%%%%%%%%%%%%%%%%%%%%%
\subsection{Azimuthal Correlations}
%%%%%%%%%%%%%%%%%%%%%%%%%%%%%%%%%%

In this section, we discuss azimuthal correlations between pairs of particles produced in pp or heavy ion collisions.  We will focus on the ellipticity and the long pseudorapidity-distance correlations or ``ridge'' phenomena later, but we will first determine the form of the probability distribution of the pairwise azimuthal angle difference within the context of our minimum-bias expansion.  Non-trivial azimuthal correlations require a non-trivial squared matrix element.  In general, the form of the terms in the squared matrix element relevant for azimuthal correlations are
\begin{align}
|{\cal M}|^2&\supset 1+\sum_{n=1}^\infty g_{n}(k^+k^-,N)\sum_{i\neq j}^N \frac{(\vec p_{\perp i}\cdot \vec p_{\perp j})^n}{Q^{2n}} \\
&\supset 1+\sum_{n=1}^\infty g_{n}(k^+k^-,N)\sum_{i\neq j}^N \frac{p_{\perp i}^n p_{\perp j}^n}{Q^{2n}} \cos(n(\phi_i-\phi_j))\nonumber\,,
\end{align}
for some coefficient functions $g_n(k^+k^-,N)$.  On the second line, we have absorbed multiplicative factors in the expansion of $(\vec p_{\perp i}\cdot \vec p_{\perp j})^n$ into a redefinition of $g_n(k^+k^-,N)$.  All other terms in the squared matrix element that are independent of azimuthal differences would just contribute to the constant ``1'' term and would therefore just be an overall normalization. With our power counting assumption that $\langle p_\perp\rangle \sim \sqrt{\langle p_\perp^2\rangle}\sim Q/N$, we replace $p_\perp \sim Q/N$ to establish the scaling with the number of observed particles $N$.  Our squared matrix element then becomes
\begin{align}
|{\cal M}|^2&\supset 1+\sum_{n=1}^\infty g_{n}(k^+k^-,N)\sum_{i\neq j}^N \frac{p_{\perp i}^n p_{\perp j}^n}{Q^{2n}} \cos(n(\phi_i-\phi_j))\\
&\sim1+\sum_{n=1}^\infty \frac{g_{n}(k^+k^-,N)}{N^{2n}}\sum_{i\neq j}^N  \cos(n(\phi_i-\phi_j))\nonumber\,.
\end{align}

The maximal scaling with $N$ of the coefficient functions $g_n(k^+k^-,N)$ can be established by demanding that the squared matrix element is non-negative.  To do this, we note that, in the $N\to\infty$ limit, transverse momentum conservation is trivially satisfied and so azimuthal angles are uncorrelated and uniformly distributed on flat phase space:
\begin{equation}
\int d\Pi_N \to \int_0^{2\pi}\prod_{i=1}^N \frac{d\phi_i}{2\pi}\,.
\end{equation}
Thus, on flat phase space, the mean of the sum over the cosine factors of the difference of azimuthal angles is 0:
\begin{equation}
\int_0^{2\pi}\prod_{i=1}^N \frac{d\phi_i}{2\pi}  \sum_{j\neq k}^N  \cos(n(\phi_j-\phi_k))=0\,.
\end{equation}
On the other hand, the central limit theorem states that the variance $\sigma^2$ of the sum of cosine factors is
\begin{align}
\sigma^2 \equiv \int_0^{2\pi}\prod_{i=1}^N \frac{d\phi_i}{2\pi}  \left[\sum_{j\neq k}^N  \cos(n(\phi_j-\phi_k))\right]^2 = N^2 \int_0^{2\pi}\prod_{i=1}^N \frac{d\phi_i}{2\pi} \, \cos^2(n(\phi_1-\phi_2)) = \frac{N^2}{2}\,,
\end{align}
as there are $N^2$ terms in the sum, in the $N\to\infty$ limit.  To ensure that the squared matrix element is non-negative, the sum over cosine factors at each value of $n$ must not be significantly negative to overwhelm the constant ``1'' term.  Therefore, we must enforce that the bulk of the possible values of the sum over cosines is less than 1:
\begin{align}
1\gtrsim \frac{g_{n}(k^+k^-,N)}{N^{2n}}\sum_{i\neq j}^N  \cos(n(\phi_i-\phi_j)) \sim \frac{g_{n}(k^+k^-,N)}{N^{2n}} \, \sigma\sim \frac{g_{n}(k^+k^-,N)}{N^{2n-1}}\,.
\end{align}
That is, the coefficient functions $g_{n}(k^+k^-,N)$ are required to scale with $N$ no greater than
\begin{align}
g_n(k^+k^-,N) \lesssim N^{2n-1}\,.
\end{align} 
This property will have important consequences for the large-$N$ predictions of azimuthal correlations.\footnote{Few general results are known about conditions for positivity of Fourier transforms.  One sufficient condition for positivity of the continuous Fourier transform of a function $u(x)$ for $x>0$ is that it is decreasing and concave-up for all $x>0$ \cite{tuck_2006,Giraud:2014sba}.  Our simple result based on scaling of terms in the Fourier series is consistent with this result.}

Continuing, we can integrate over the smeared phase space to establish the probability distribution for the pairwise azimuthal angle difference, $\Delta\phi$.  We have
\begin{align}\label{eq:azidist}
p(\Delta \phi) &\sim\frac{1}{Q^2} \int_0^Q dk^+ \int_0^Q dk^- \, f(k^+k^-)\, \int d\Pi_N |{\cal M}|^2\, \delta(\Delta\phi -(\phi_1-\phi_2))\\
&\sim \frac{1}{Q^2} \int_0^Q dk^+ \int_0^Q dk^- \, f(k^+k^-) \int_0^{2\pi} \prod_{i=1}^N \frac{d\phi_i}{2\pi}\, \delta(\Delta\phi -(\phi_1-\phi_2))\nonumber\\
&
\hspace{1cm}\times\left(
1+\sum_{n=1}^\infty \frac{g_{n}(k^+k^-,N)}{N^{2n}}\sum_{i\neq j}^N \cos(n(\phi_i-\phi_j))
\right)\nonumber\\
&=\frac{1}{2\pi}+\frac{1}{\pi}\sum_{n=1}^\infty \frac{d_n(N)}{N^{2n}}\, \cos(n\, \Delta\phi)\,.\nonumber
\end{align}
Here, we have used the permutation symmetry of particles to just define $\Delta \phi$ as the difference of the azimuthal angles of particles 1 and 2.  The Fourier coefficients $d_n(N)$ are defined to be
\begin{equation}\label{eq:fcoeffsazi}
d_n(N) =  \frac{1}{2Q^2} \int_0^Q dk^+ \int_0^Q dk^- \, f(k^+k^-)\, g_{n}(k^+k^-,N)\,.
\end{equation}
Note that the maximal $N$ scaling of the coefficients $d_n(N)$ is inherited from the form of $g_n(k^+k^-,N)$ established above by positivity; that is,
\begin{equation}
d_n(N) \lesssim N^{2n-1}\,.
\end{equation}
This scaling means that the Fourier coefficients at each $n$ necessarily vanish in the $N\to\infty$ limit:
\begin{equation}\label{eq:fourcoefflimit}
\lim_{N\to\infty}\frac{d_n(N)}{N^{2n}} = 0\,.
\end{equation}
The distribution on the final line of \Eq{eq:azidist} is normalized on $\Delta\phi \in[0,2\pi)$.\footnote{Recall that we are working at fixed center-of-mass energy $Q$.  It is known that azimuthal correlations have a finite, non-zero value at fixed centrality as both $Q$ and $N$ increase (see, e.g., \cite{Aamodt:2010pa}).  This is not inconsistent with this analysis because the Fourier coefficients defined in \Eq{eq:fcoeffsazi} have implicit dependence on the center-of-mass energy $Q$ and other mass scales in the system, like the pion mass $m_\pi$.}

The particular scaling of the coefficient $d_1(N)$ with $N$ can also be established by momentum conservation.  Transverse momentum conservation states that
\begin{align}
0=\left(
\sum_{i=1}^N \vec p_{\perp i}
\right)^2=\sum_{i=1}^N p_{\perp i}^2+\sum_{i\neq j}^N p_{\perp i}p_{\perp j}\cos(\phi_i-\phi_j)\,.
\end{align}
Now, with $p_{\perp}\sim Q/N$, this relationship implies the scaling
\begin{equation}
\frac{1}{N^2}\sum_{i\neq j}^N \cos(\phi_i-\phi_j) \sim -\frac{1}{N}\,.
\end{equation}
With our ergodic assumption, the term on the left is just the mean value of $\cos(\Delta\phi)$ in the large-$N$ limit and so
\begin{equation}
\langle \cos(\Delta \phi)\rangle \sim -\frac{1}{N}\,.
\end{equation}
This mean value can also be calculated from the probability distribution $p(\Delta\phi)$.  We have
\begin{equation}
\langle \cos(\Delta \phi)\rangle = \int_0^{2\pi} d\Delta\phi\, p(\Delta\phi)\, \cos(\Delta\phi) = \frac{d_1(N)}{N^{2}}\,,
\end{equation}
from \Eq{eq:azidist}.  If this is to scale like $-1/N$, the coefficient $d_1(N)$ must scale like
\begin{equation}
d_1(N) \sim -N = -N^{2\cdot 1-1}\,,
\end{equation}
exactly as predicted from positivity.\footnote{Yet another way to prove this scaling with $N$ follows from demanding that the squared matrix element is finite in the $N\to\infty$ limit.}  Then, the distribution of the azimuthal angle difference can be expressed as
\begin{align}\label{eq:delphidistfinal}
p(\Delta\phi)  =\frac{1}{2\pi}-\frac{d_1}{\pi}\frac{1}{N}\cos(\Delta\phi) +\frac{1}{\pi}\sum_{n=2}^\infty \frac{d_n(N)}{N^{2n}}\, \cos(n\, \Delta\phi)\,,
\end{align}
where now $d_1>0$ is some constant value in the $N\to\infty$ limit.  A similar relationship for the $n=1$ coefficient of the Fourier expansion using momentum conservation was established in \Ref{Luzum:2010fb}, but to the best of our knowledge, the scaling with the number of particles $N$ is novel.

\subsubsection{Ellipticity}

Especially in collisions of heavy ions, azimuthal correlations amongst particles are used as evidence for collective flow phenomena and the production of exotic states of QCD matter.  The first non-trivial azimuthal correlation is referred to as elliptic flow and quantifies particle correlations with respect to the reaction plane of the collision, the plane about which particle production is maximized in the plane and minimized orthogonal to it.  As a proxy for the elliptic flow, a pairwise azimuthal correlation moment is often measured instead.  Here, we will focus on one such moment, the average over all $N$ particles in an event denoted as $\langle 2\rangle$ and defined to be
\begin{align}
\langle 2\rangle=\frac{1}{N^2}\sum_{j,k=1}^N e^{2i(\phi_j-\phi_k)}&=\frac{1}{N}+\frac{1}{N^2}\sum_{j\neq k}^N\cos\left(2(\phi_j-\phi_k)\right)\,,
\end{align}
where $\phi_j$ is the azimuthal angle of particle $j$.  As above, the ergodic assumption establishes that the sum over cosine factors is just the mean value of cosine of the azimuthal angle difference in the large $N$ limit, and so
\begin{equation}\label{eq:azcorrdef}
\langle 2\rangle=\frac{1}{N}+\langle\cos\left(2\Delta \phi\right)\rangle\,.
\end{equation}
When further averaged over an ensemble of events, this is often denoted as $\langle\langle 2\rangle\rangle$ or $c_2(2)$.  The notation $c_2(2)$ is common in the literature with the parenthetical 2 referencing the pairwise azimuthal correlation and the subscript 2 referencing the 2 in the argument of cosine in \Eq{eq:azcorrdef}.

Azimuthal correlations have been extensively measured in heavy ion collisions for the past few decades, in experiments at ATLAS \cite{Aad:2015gqa,Aad:2014eoa,ATLAS:2011ah}, CMS \cite{Sirunyan:2017fts,Chatrchyan:2012ta,Chatrchyan:2013nka,Sirunyan:2017uyl,Sirunyan:2017gyb,Sirunyan:2017pan,CMS:2013bza,Chatrchyan:2013kba,Chatrchyan:2012vqa,Chatrchyan:2012xq,Chatrchyan:2012wg}, and ALICE \cite{Acharya:2019vdf,Adam:2016izf,Abelev:2014mda,Adam:2015eta,Aamodt:2010pa} at the LHC and at STAR \cite{Adamczyk:2012ku,Agakishiev:2011id,Abelev:2008ae,Adler:2002pu,Ackermann:2000tr,Adam:2019woz,Adamczyk:2017ird,Adamczyk:2015obl,Abdelwahab:2014sge,Adams:2004bi,Adams:2004wz} and PHENIX \cite{Adare:2018zkb,Adare:2010ux} at RHIC.  However, outside of an explicit hydrodynamic model, measurements of azimuthal correlations are typically challenging to interpret.  Azimuthal correlations are often binned in unphysical or unobservable quantities that are fit from models of the ion collisions, like the number of nucleons participating in the collision or the centrality.\footnote{Experimentally, the centrality is measurable as it is defined as a quantile of the multiplicity distribution for a given collision event data set.  However, the centrality is not defined for an individual event in isolation and ``centrality'' connotes the overlap of the ion nuclei in collision, which is not measurable.}  Few of the listed references plot azimuthal correlations as a function of directly observable quantities, such as the number of charged particles in the event \cite{Aad:2015gqa,Chatrchyan:2013nka,Sirunyan:2017uyl,Acharya:2019vdf,Abelev:2014mda,Adamczyk:2015obl}.  In what follows we will show how  the features of the observables can in fact be elucidated purely in terms of observable quantities, by appealing to the expansion of the minimum bias cross section derived above.

As mentioned earlier, our expansion of the minimum bias cross section is defined for a fixed collision energy $Q$ and fixed number $N$ of observed particles.  With our current formulation, we are not able to predict the dependence of the azimuthal correlations either as a function of $Q$ or $N$.  Thus, we will focus on predictions for fixed $Q$ binned in the number of observed particles.  Nevertheless, we will be able to make a number of concrete predictions.  For fixed $N$ and $Q$, the average azimuthal correlation $c_2(2)$ in terms of our expansion of minimum bias established in \Eq{eq:delphidistfinal} is
\begin{align}
c_2(2) &=\frac{1}{N}+\int_0^{2\pi}d\Delta\phi\, p(\Delta\phi)\, \cos(2\Delta\phi) = \frac{1}{N} + \frac{d_2(N)}{N^4}\,.
\end{align}
This form immediately identifies two distinct contributions.  First, the explicit $1/N$ term is completely independent of any of the dynamics of the collision (i.e., the squared matrix element and smearing).  Such a contribution is sometimes called ``non-flow'' in the literature.  The second term, by contrast, is only non-zero if there are non-trivial azimuthal correlations; otherwise, the integral over phase space of the sinusoidal function vanishes.  Such a contribution is sometimes called ``flow''.  Whatever its short-distance interpretation, we can make concrete predictions of the flow contribution within our expansion.

First, the azimuthal correlation $c_2(2)$ vanishes in the $N\to\infty$ limit.  This follows directly from the results established in \Eq{eq:fourcoefflimit}.  Additionally, this large-$N$ limit is also the limit of small centrality (the quantile of highest multiplicity in an ensemble of collision events).  Thus, azimuthal correlations should also vanish in the limit of small centrality.  In the nucleus-overlap model of heavy ion collisions, it is also predicted that azimuthal correlations vanish in the low centrality limit because a head-on, perfectly overlapping nucleus collision is completely rotationally symmetric and has no preferred particle production axis.  However, we stress that this prediction in our formulation makes no reference to the unobservable nucleus overlap and relies instead exclusively on our power counting of the relevant observable quantities.

\begin{figure}[t]
\begin{center}
\includegraphics[width=9cm]{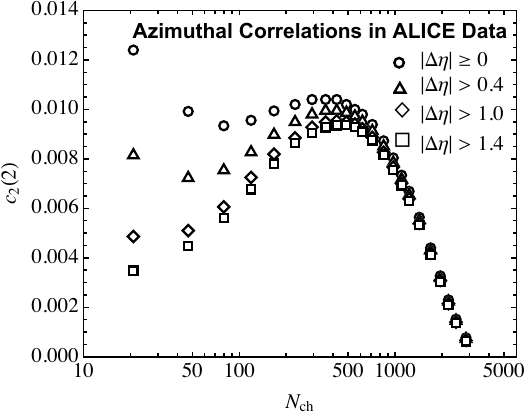}
\caption{Data of the azimuthal correlation $c_2(2)$ as a function of charged particle multiplicity $N_\text{ch}$ from the ALICE experiment \cite{Abelev:2014mda}.  Lead ions are collided with a nucleon-nucleon center of mass of $\sqrt{s_\text{NN}} = 2.76$ TeV.  Only charged particles with pseudorapidity $|\eta|<1$ and transverse momentum $0.2 < p_\perp <3.0$ GeV contribute.  The open circle data include all charged particles, and the open triangle, diamond, and square data only include those pairs of charged particles with a pseudorapidity difference of at least 0.4, 1.0, and 1.4, respectively.
\label{fig:ellipNch}}
\end{center}
\end{figure}

This prediction is observed in data from heavy ion collisions.  In \Fig{fig:ellipNch}, we plot data of the azimuthal correlation $c_2(2)$ from PbPb collisions at the ALICE experiment, as a function of the number of charged particles $N_\text{ch}$ in the pseudorapidity range $|\eta| < 1.0$ and transverse momentum in the range $0.2 < p_\perp < 3.0$ GeV \cite{Abelev:2014mda}.  Only charged particles are used to compute $c_2(2)$, but as discussed earlier, as long as the number of particles is large, we still expect our expansion to accurately describe these events.  For significantly large $N_\text{ch}$ the azimuthal correlation decreases toward 0, as our expansion predicts.  Though we do not show a plot here, data of the azimuthal correlation binned in centrality also vanishes as centrality decreases (see, e.g., \Ref{Adam:2016izf}), corresponding to the highest multiplicity events in an ensemble.

The plot presented in \Fig{fig:ellipNch} also illustrates another feature described in our expansion.  The four data sets in that plot correspond to the azimuthal correlation of pairs of particles with a minimal pseudorapidity difference, $|\Delta\eta|$.  As the pairwise pseudorapidity difference increases, the azimuthal correlation at small multiplicity $N_\text{ch}$ significantly decreases.  At small multiplicities, momentum conservation induces significant correlations between pairs of particles that is imprinted on $c_2(2)$, especially when further restricted to long-range pseudorapidity correlations.  However, at large values of $N_\text{ch}$ the data with different pseudorapidity differences converge onto a universal curve.  This is expected within our expansion.  At large $N_\text{ch}$, correlations induced by momentum conservation are eliminated.  Further, the pseudorapidity distribution of particles is quite flat over the acceptance region of the detector, and we argued that the transverse momentum distribution of particles effectively independent of a pseudorapidity cut.  So, for sufficiently large number of particles, there is nothing special about the particular pseudorapidity of a particle and its correlation to other particles.  Therefore, in the large $N_\text{ch}$ limit, neighboring particles and far away particles in pseudorapidity will exhibit the same azimuthal correlations.

%%%%%%%%%%%%%%%%%%%%%%%%%%%%%%%%%%
\subsubsection{Two-Particle Angular Correlations; the ``Ridge''}
%%%%%%%%%%%%%%%%%%%%%%%%%%%%%%%%%%

A more general analysis of two-particle correlations can be studied through measurement of the cross section differential in both pairwise azimuthal differences $\Delta\phi$ as well as pairwise pseudorapidity differences $\Delta\eta$.  Such double differential cross sections have been measured at RHIC \cite{Abelev:2009af,Alver:2009id} and other experiments before the LHC, but became exciting evidence of potential collective phenomena in pp collisions in measurements at CMS \cite{Khachatryan:2010gv} in 2010.  Since then, CMS \cite{Chatrchyan:2011eka,Chatrchyan:2012wg,Chatrchyan:2013nka,CMS:2013bza,Khachatryan:2015lva,Khachatryan:2016txc}, ATLAS \cite{ATLAS:2012ap,Aad:2015gqa,Aaboud:2016yar,ATLAS:2021jhn}, and ALICE \cite{Aamodt:2011by,Adam:2015gda} have measured pairwise particle correlations in pp and PbPb ion collisions.  In heavy ion collisions, it is well-established that there are long range in $\Delta \eta$ correlations, which is interpreted as collective phenomena due to the high-density medium produced in collision.  Extending out to $\Delta\eta\sim 4$, the distribution of azimuthal correlations in $\Delta\phi$ is nearly independent of $\Delta\eta$ and exhibit an over-density in the distribution around $\Delta\phi = 0$, the so-called ``ridge''.  A long distance factorization of the $\Delta\eta$ and $\Delta \phi$ distributions and the ridge are challenging to observe in pp collisions because of the overwhelming dominance of short-distance correlations, or jet phenomena.  However, with significantly restrictive cuts at high particle multiplicity, the CMS analysis from 2010 and later measurements identified a small, but non-negligible ridge in pp collisions suggesting that a dense QCD medium was produced.

In this section, we will analyze these angular correlations and the ridge from the perspective of our expansion of high multiplicity, minimum bias events.  First, the quantity that is typically measured and plotted to identify the ridge is the cross section ratio
\begin{align}
\frac{d^2 N^\text{pair}}{d\Delta\eta \, d\Delta\phi} = B(0,0)\times \frac{S(\Delta\eta,\Delta\phi)}{B(\Delta\eta,\Delta\phi)}\,.
\end{align}
Here, $N^\text{pair}$ is the number of pairs of particles from which their pseudorapidity difference $\Delta\eta$ and azimuthal angle difference $\Delta \phi$ are measured.  $S(\Delta\eta,\Delta\phi)$ is the double differential cross section for ``signal'' pairs of particles drawn from the same event.  To eliminate trivial correlations to some extent, this is divided by the ``background'' distribution $B(\Delta\eta,\Delta\phi)$ which consists of pairs of particles drawn from different events.  Thus, the paired particles that contribute to $B(\Delta\eta,\Delta\phi)$ are truly uncorrelated.  Finally, $B(0,0)$ is a convenient normalization factor.

Starting with the signal distribution $S(\Delta\eta,\Delta\phi)$, it is defined to be
\begin{align}
S(\Delta\eta,\Delta\phi) = \frac{d^2N^\text{same}}{d\Delta\eta \, d\Delta\phi}\,,
\end{align}
where $N^\text{same}$ is the number of pairs of particles drawn from the same event.  In the limit that the number of detected particles in the event $N\to\infty$, correlations between pairs of particles due to momentum conservation are eliminated.  Further, we expect that the pseudorapidity and azimuthal angle differences are uncorrelated, or only very weakly correlated, because the squared matrix element approaches a constant in this limit.  Any such correlations have to come  from terms in the expansion in Eq.~\eqref{eq:expansion0} of at least  order $Q^{-4}$, and so the signal distribution factorizes to leading power in $1/N$ as
\begin{align}
S(\Delta\eta,\Delta\phi) = \frac{d^2N^\text{same}}{d\Delta\eta \, d\Delta\phi} \sim \frac{dN^\text{same}}{d\Delta\eta}\times\frac{1}{N^\text{same}}\frac{dN^\text{same}}{d\Delta\phi} = \frac{dN^\text{pair}}{d\Delta\eta}\, p(\Delta\phi)\,.
\end{align}
Here, $p(\Delta\phi)$ is the probability distribution for azimuthal angle differences established in \Eq{eq:delphidistfinal}, and $dN^\text{pair}/d\Delta\eta$ is the pairwise $\Delta\eta$ distribution.

Continuing to the background distribution $B(\Delta\eta,\Delta\phi)$, we expect no correlation between $\Delta\eta$ and $\Delta\phi$ for pairs of particles from different events, so this distribution would naturally factorize.  Further, the azimuthal difference should be flat because the experiment is symmetric about the colliding beams and each event spontaneously breaks this rotational symmetry, but independent of all other events.  We also expect that the pseudorapidity difference distribution of the background is identical to that of signal because in the large $N$ limit, correlations between pairs of particles in the same event vanish.  Thus, the pseudorapidity difference distribution for same-event or distinct-event pairs is just determined by drawing from the single particle pseudorapidity distribution twice, to leading power in $1/N$.  That is, the background distribution factorizes to leading power in $1/N$ as
\begin{align}
B(\Delta\eta,\Delta\phi) = \frac{d^2N^\text{mix}}{d\Delta\eta \, d\Delta\phi} \sim \frac{dN^\text{mix}}{d\Delta\eta}\times\frac{1}{N^\text{mix}}\frac{dN^\text{mix}}{d\Delta\phi} = \frac{1}{2\pi}\frac{dN^\text{pair}}{d\Delta\eta}\,,
\end{align}
where $N^\text{mix}$ is the number of pairs of particles from distinct or ``mixed'' events.

With these assumptions, our prediction for the double differential pairwise correlation distribution is independent of $\Delta\eta$ and just determined by the azimuthal difference distribution:
\begin{align}
\frac{d^2 N^\text{pair}}{d\Delta\eta \, d\Delta\phi} \propto p(\Delta\phi)\,.
\end{align}
The large-$N$ expansion of the minimum bias cross section immediately implies that there are long range in $\Delta\eta$ correlations, through the interpretation that the $\Delta\phi$ distribution at disparate values of $\Delta\eta$ are, to first order, the same.  While these long-range correlations are observed in data and are a requirement for existence of a ridge, it doesn't by itself predict an over-density of pairs of particles with $\Delta\phi = 0$.  That requires the distribution $p(\Delta\phi)$ to have a maximum at $\Delta\phi = 0$.  One can perform an analysis of the coefficients of the Fourier expansion of $p(\Delta\phi)$, and compare to data.  Recall that our large-$N$ expansion of minimum bias predicts the form
\begin{equation}
p(\Delta\phi)  =\frac{1}{2\pi}-\frac{d_1}{\pi}\frac{1}{N}\cos(\Delta\phi) +\frac{1}{\pi}\sum_{n=2}^\infty \frac{d_n(N)}{N^{2n}}\, \cos(n\, \Delta\phi)\,,
\end{equation}
where $d_1>0$ and the $d_n(N)$ coefficients scale at worst like $N^{2n-1}$ in the large $N$ limit.  Additionally, from ellipticity, we know that $d_2(N)>0$, at least for minimum bias events in PbPb collisions. 

For existence of a ridge, we would additionally need that 
\begin{align}
&\left.\frac{dp(\Delta\phi)}{d\Delta\phi}\right|_{\Delta\phi = 0}=0\,,
&\left.\frac{d^2p(\Delta\phi)}{d\Delta\phi^2}\right|_{\Delta\phi = 0}<0\,.
\end{align}
The vanishing of the first derivative at $\Delta\phi=0$ is guaranteed by the symmetries of minimum bias collisions.  The concave down requirement constrains the Fourier coefficients, but without a short-distance theory from which to predict the $d_n(N)$ values, we can't say much more within the context of our minimum bias expansion.  However, the fact that the distribution $p(\Delta\phi)$ is physical highly constrains the coefficients $d_n(N)$.  Assuming that a physical distribution is infinitely differentiable and the Fourier series of every derivative is absolutely convergent, the $d_n(N)$ must vanish as $n\to\infty$ faster than any power of $n$.  For example, the Fourier series could actually just be finite for which there is a maximum $n$ at which $d_n(N)$ is non-zero, or $d_n(N)$ might vanish exponentially fast, like $d_n(N)\sim e^{-n}$.  Infinite differentiability implies that there are only a finite number of terms in the Fourier expansion that can have an appreciable effect on the first and second derivatives of $p(\Delta \phi)$ at $\Delta \phi=0$.  This seems to be borne out in data, e.g., figure 6 of \Ref{Chatrchyan:2011eka}, in which the $d_2(N)$, $d_3(N)$, and $d_4(N)$ coefficients would be positive, but $d_5(N)$ appears to be vanishingly small.  Of course, with any finite dataset, measuring higher Fourier modes is problematic, but the trend is consistent with expectations from differentiability.

%%%%%%%%%%%%%%%%%%%%%%%%%%%%%%%%%%
\section{Min-Bias in Other Colliders}\label{sec:other}
%%%%%%%%%%%%%%%%%%%%%%%%%%%%%%%%%%

While our focus in this paper has been the application of the large-$N$ expansion of minimum bias events to understand identical hadron or identical heavy ion collisions, the power counting and symmetries can be appropriately modified to describe nearly any collision environment.  In this section, we will just briefly describe extension to minimum bias at other colliders.  Minimum bias in electron-positron collisions will be significantly different that we will provide more details about its symmetries and predictions.

%%%%%%%%%%%%%%%%%%%%%%%%%%%%%%%%%%
\subsection{pA Collisions and Electron-Ion Collisions}
%%%%%%%%%%%%%%%%%%%%%%%%%%%%%%%%%%

For different types of colliders, we must modify our power counting and symmetry assumptions as established in \Sec{sec:ppmin}.  Here, we will just discuss how the power counting and symmetries are modified for proton-ion (pA) and electron-ion collisions, leaving a detailed analysis of unique predictions from our minimum bias expansion in those environments to future work.  First, for pA collisions, as the initial state still consists of hadronic matter, there are still unmeasurable beam regions in which all we can determine is the total momentum lost down the beam.  Thus, all power counting assumptions for minimum bias in pA collisions remain identical to that of pp/AA collisions.  The only change is to the symmetries, in which pA collisions lack the beam reflection $\eta\to -\eta$ symmetry of pp/AA collisions.  As a consequence, the expansion of the corresponding matrix element is not reflection symmetric about the beams and so can have terms that consist of the hyperbolic sine of pseudorapidity differences, as well as hyperbolic cosine.  For example, the squared matrix element for the $N$ detected particles of pA collisions can now have terms of the form
\begin{equation}
|{\cal M}|^2\supset \sum_{i\neq j}^N p_{\perp i}p_{\perp j}\sinh(\eta_i-\eta_j)\,.
\end{equation}

For electron-ion colliders, the power counting and symmetries are even more different.  Now, because the electron is not hadronic, there is only one unmeasurable beam region, in the direction of the momentum of the initial ion.  This isn't necessarily to say that there is perfect experimental resolution in the direction of the electron's momentum, it is instead that because the electron is not hadronic, its direction of momentum is not special for establishing beam remnants that take away an order-1 energy fraction at very high pseudorapidity.  As there is only one unmeasurable beam region, we only smear over one component of lightcone momentum, $k^+$, say.  This also means that the pseudorapidity translation symmetry $\eta\to \eta+\Delta\eta$ is broken because we can effectively measure all particles at arbitrary pseudorapidities in the direction of the electron's momentum.  Additionally, as the colliding particles are not identical, there is no beam reflection symmetry, $\eta\to-\eta$.  Thus in this case, the cross section for minimum bias events takes the form:
\begin{align}
\sigma &\sim \frac{1}{Q}\int_0^Q dk^+\,f(k^+) \int d\Pi_N\, |{\cal M}|^2\\
&
\hspace{2cm}\times\delta\left(
k^+ - \sum_{i=1}^N p_{\perp i}e^{-\eta_i}
\right)\, \delta\left(
Q - \sum_{i=1}^N p_{\perp i}e^{\eta_i}
\right)\, \delta^{(2)}\left(
\sum_{i=1}^N \vec p_{\perp i}
\right)\nonumber\,.
\end{align}
Here, $f(k^+)$ is some function purely of the total $k^+$ momentum lost down the beampipe in the direction of the initial ion's momentum and $|{\cal M}|^2$ is the squared matrix element of the $N$ detected particles.  Decreasing the constraining symmetries makes the expansion of the squared matrix element significantly more involved, so we leave predictions for electron-ion collisions to future work.

%%%%%%%%%%%%%%%%%%%%%%%%%%%%%%%%%%
\subsection{Electron-Positron Collisions}
%%%%%%%%%%%%%%%%%%%%%%%%%%%%%%%%%%

At high energies in electron-positron collisions, the total hadronic cross section is dominantly due to collisions of emitted, collinear photons.  In these events, the electron and positron are generally only scattered at low angle and are therefore lost down the beam.  Because of the soft singularity of QED, the emitted photons typically have a small energy compared to the center-of-mass collision energy, and so the observed energy in the detector is small.  This configuration of minimum bias $e^+e^-$ collision events therefore is very similar in experimental signature as minimum bias events in hadron collisions.  Specifically, in experimental analyses of these secondary $\gamma\gamma\to$ hadrons events, they are identified by an anti-tag of energetic electron and positrons \cite{L3:1997adt,OPAL:1999pnw}, requiring that the colliding particles are lost down the beam, and a limit on the total energy to remove annihilation events.  With these considerations, the minimum bias events in $e^+e^-$ collisions would enjoy the same power counting and symmetries as described in \Sec{sec:ppmin} and the expansion of the cross section would also follow from that presented there.

Here, 
%however, we will not study these continuum photon minimum bias events from $e^+e^-$ collisions.  
instead, we will analyze the structure of events from $e^+e^-$ collisions at or near a resonance, like the $Z$ pole, $m_Z$.  At a center-of-mass energy near $m_Z$, the cross section is dominated by the $Z$ boson resonance.  This configuration of final state particles is very different than minimum bias in hadron collisions, or the effective $\gamma\gamma$ collisions away from the resonance.  As such, the power counting and symmetries of these resonance events are distinct, and produce different predictions for the structure of events and the correlations between particles. We will take these conditions to  define events that we term `resonance-decay minimum bias'.

The description of resonance-deay, hadronic minimum bias events in electron-positron ($e^+e^-$) collisions is sufficiently constraining and simple that we will explicitly construct the large-$N$ expansion and make some predictions.  First and foremost, because electrons are color-singlet, fundamental particles, there are no unmeasurable beam regions whatsoever, so (in principle) all final state particles can be detected.  Consequently, because the initial state is the hadronic vacuum, we assume full Lorentz invariance of the final state, so natural phase space coordinates are the energies of particles and their location on the celestial sphere.  Further, as with our pp/AA collision analysis, we assume that only momenta are measured; no information on electric charge, etc., is recorded for particles.

With these assumptions, the power counting for resonance-decay hadronic minimum bias events in $e^+e^-$ collisions are:
\begin{enumerate}

\item All $N$ particles produced  can in principle be detected.  Only their four-momenta are measured; no charge information is determined.

\item  We assume that the mass of the particles is irrelevant and so detected particle energy $E$ is parametrically larger than the QCD scale or pion mass, $E \gg m_\pi$.

\item The number of detected particles $N$ is large: $N\gg1$.

\item We assume that the mean energy of the detected particles is representative of all particles' energies and so the mean and the root mean square energies are comparable: $\langle E\rangle\sim \sqrt{\langle E^2\rangle} \sim Q/N$.
\end{enumerate}
In contrast to the hadronic cross section due to collisions of collinear photons, there is no assumption that order 1 of the collision energy is lost down the beam.
These power counting assumptions then imply that these minimum bias events enjoy the following symmetries:
\begin{enumerate}

\item Complete Lorentz invariance O$(3,1)$ and parity symmetry of the final state,

\item reflection of the electron-positron beams because no charge information is measured, and

\item $S_N$ permutation symmetry in all $N$ detected particles.
\end{enumerate}
The reflection symmetry of the beams is already accounted for with the O(3,1) Lorentz invariance, but we mention it explicitly because it would not be true if charge information were retained.

With these symmetries, the cross section for minimum bias events can just be expressed in the textbook form of Fermi's Golden Rule with four-momentum conservation in the center-of-mass frame:
\begin{align}
\sigma &\sim \int \Pi_N\, |{\cal M}|^2\sim \int \prod_{i=1}^N\frac{d^3 p_i}{2E_i}\, |{\cal M}|^2\, \delta^{(4)}\left(
Q-\sum_{i=1}^N p_i
\right)\\
&\sim \int \prod_{i=1}^N \left[E_i\, d E_i\, d\cos\theta_i\, \frac{d\phi_i}{2\pi}\right]\, |{\cal M}|^2\, \delta\left(
Q-\sum_{i=1}^N E_i
\right)\, \delta\left(
\sum_{i=1}^N E_i\cos\theta_i
\right)\,\delta^{(2)}\left(
\sum_{i=1}^N \vec p_{\perp i}
\right)\,,
\nonumber
\end{align}
where $|{\cal M}|^2$ is the Lorentz-invariant squared matrix element for $N$ final state particles, $E_i$ is the energy of particle $i$, and $\theta_i$ and $\phi_i$ are its polar and azimuthal angle coordinates on the celestial sphere.  As it is fully Lorentz invariant, the squared matrix element can be expanded in symmetric polynomials of the Mandelstam invariants $s_{ij} = 2p_i\cdot p_j$, for two particles' four-momenta $p_i$ and $p_j$.  Up through mass dimension 4 terms, the squared matrix element can be expanded in symmetric polynomials as:
\begin{align}
|{\cal M}|^2 &=1+\frac{c_1^{(4)}}{Q^4}\sum_{i\neq j\neq k\neq l}^N s_{ij}s_{kl}+\frac{c_2^{(4)}}{Q^4}\sum_{i\neq j\neq k}^Ns_{ij}s_{ik}+\frac{c_3^{(4)}}{Q^4}\sum_{i\neq j}^Ns_{ij}^2+{\cal O}(Q^{-6})
\end{align}
The $c_i^{(n)}$ factors are constants that may in general have dependence on the number of particles $N$.  The only constraint we impose on them is that in the $N\to\infty$ limit, their scaling with $N$ produces a finite contribution to the squared matrix element and that they are not sufficiently negative to make the squared matrix element negative.  For example, assuming that the energy of particle $i$ scales like $E_i\sim Q/N$ and so the second non-trivial term scales with $N$ as
\begin{equation}
\sum_{i\neq j\neq k}^Ns_{ij}s_{ik} \sim N^3\frac{Q^4}{N^4}\sim \frac{Q^4}{N}\,,
\end{equation}
because there are $N^3$ terms in the sum in the $N\to\infty$ limit.  Therefore, for finiteness and positivity, the coefficient $c_2^{(4)}$ is bounded as
\begin{equation}
-N \lesssim c_2^{(4)}\lesssim N\,.
\end{equation}

As discussed in the expansion of the squared matrix element for pp/AA collisions, in the $N\to\infty$ limit, the various terms in the squared matrix element relax to their ensemble mean with 0 variance, with our assumption of ergodicity and the central limit theorem.  So, in the $N\to\infty$ limit, our minimum bias power counting implies that the matrix element is just a constant, corresponding to $N$ free final state particles only constrained by momentum conservation.

%%%%%%%%%%%%%%%%%%%%%%%%%%%%%%%%%%
\subsubsection{Suppression of a Ridge in Resonance Decay Min Bias}
%%%%%%%%%%%%%%%%%%%%%%%%%%%%%%%%%%

Within this hadronic resonance decay minimum bias expansion for $e^+e^-$ collision events, we will concretely demonstrate that this predicts a strong suppression of possible ridge phenomena in azimuthal correlations of pairs of particles.  A ridge in two-particle correlations was searched for in $e^+e^-$ collisions at the $Z$ pole from archived ALEPH data recently \cite{Badea:2019vey}, but no evidence for such a ridge was found.  Lack of azimuthal correlations over a long distance in pseudorapidity in $e^+e^-$ collisions is perhaps not surprising because the initial state is the QCD vacuum and there are no special directions in which particles are produced at arbitrarily high pseudorapidities.  Nevertheless, it is satisfying to see that the power counting and symmetries of $e^+e^-$ collisions immediately predict that such correlations, if they do exist, are very suppressed in the large-$N$ limit.

To demonstrate this suppression of azimuthal correlations, we will focus on terms in the squared matrix element that directly correlate pairs of particles:
\begin{align}
|{\cal M}|^2 &\supset 1+\sum_{n=2}^\infty c_n\sum_{i\neq j}^N \frac{s_{ij}^n}{Q^{2n}} = 1+\sum_{n=2}^\infty \frac{2^nc_n}{Q^{2n}}\sum_{i\neq j}^N p_{\perp i}^np_{\perp j}^n\left(
\cosh \Delta \eta_{ij} - \cos\Delta \phi_{ij}
\right)^n\,.
\end{align}
Here, the $c_n$ are some constants that possibly depend on $N$ but are constrained by finiteness and on the right, we have expanded the Mandelstam invariant $s_{ij}$ in terms of cylindrical detector coordinates.  Now, we would like to simplify this expression, focusing on the scaling with $N$ and the azimuthal correlations explicitly, so we will make some replacements.  First, the transverse momentum $p_\perp = E\sin\theta$ and on flat phase space, most particles are located around the ``equator'' of the celestial sphere, where $\theta = \pi/2$.  So, as $E\sim Q/N$, so too does $p_\perp \sim Q/N$ in $e^+e^-$ collisions.  Further, because most particles lie near the equator, the hyperbolic cosine of the pairwise pseudorapidity difference is close to 1, so we can take the scaling $\cosh \Delta \eta \sim 1$.  With these scaling assumptions, the squared matrix element takes the form:
\begin{align}
|{\cal M}|^2 &\supset  1+\sum_{n=2}^\infty \frac{2^nc_n}{Q^{2n}}\sum_{i\neq j}^N p_{\perp i}^np_{\perp j}^n\left(
\cosh \Delta \eta_{ij} - \cos\Delta \phi_{ij}
\right)^n\\
&\sim1+\sum_{n=2}^\infty \frac{2^nc_n}{N^{2n}}\sum_{i\neq j}^N \left(
1 - \cos\Delta \phi_{ij}
\right)^n\nonumber\,.
\end{align}

Now, with this squared matrix element, we can determine the probability distribution of the pairwise azimuthal angle difference $\Delta\phi$, in the large-$N$ limit.  As in our analysis of pp/AA collisions, correlations from momentum conservation become trivial in the $N\to\infty$ limit, and the probability distribution can be calculated from
\begin{align}
p(\Delta\phi)&\propto \int_0^{2\pi}\prod_{k=1}^N\frac{d\phi_k}{2\pi}\, |{\cal M}|^2 \, \delta(\Delta \phi-(\phi_1-\phi_2))\\
&= \int_0^{2\pi}\prod_{k=1}^N\frac{d\phi_k}{2\pi}\,\left(
1+\sum_{n=2}^\infty \frac{2^nc_n}{N^{2n}}\sum_{i\neq j}^N \left(
1 - \cos\Delta \phi_{ij}
\right)^n
\right)\, \delta(\Delta \phi-(\phi_1-\phi_2))\nonumber\\
&=\frac{1}{2\pi}+\frac{1}{\pi}\sum_{n=2}^\infty \frac{2^nc_n}{N^{2n}}\left[
\frac{2^{n-1}\Gamma\left(\frac{1}{2}+n\right)}{\sqrt{\pi}\Gamma(1+n)} \,(N+1)(N-2)+\left(
1 - \cos\Delta \phi
\right)^n
\right]
\nonumber
\end{align}
We have used permutation symmetry to define $\Delta\phi = \phi_1-\phi_2$, and the result in the final line is the exact integral over azimuthal angles of the second line.  In this final form, it is obvious that any azimuthal dependence is suppressed by a relative factor of $N^2$ in the large-$N$ limit.  This explicitly follows from Lorentz invariance of the final state, enforcing that the transverse and longitudinal momentum components to the beam must appear in the expansion of the squared matrix element with the same coefficient.  

On the other hand, the minimum bias expansion for the hadronic events arising from secondary $\gamma \gamma$ collisions discussed at the beginning of this section predicts no such suppression of a ridge, so it would be of interest to redo the analysis of \cite{Badea:2019vey} away from the $Z$ pole.

%%%%%%%%%%%%%%%%%%%%%%%%%%%%%%%%%%
\section{Outlook}\label{sec:concs}
%%%%%%%%%%%%%%%%%%%%%%%%%%%%%%%%%%

In this paper, we elucidated a power counting scheme to describe minimum bias events at colliders. The main tenets of this approach are a principle of particle ergodicity---each particle is representative of any other particle and averages over particles in an event are equivalent to averages of events over an ensemble---and the assumption of large particle multiplicity, $N$, in an event which provides the expansion parameter $1/N$. Under these conditions, the {\it variance} of the QCD squared matrix element vanishes as $1/N$, and as $N\to \infty$, this effectively reduces all physical observables to their flat phase space limits, with a smearing factor in p or A collisions to account for beam losses.

We showed that the above assumptions do indeed lead to a remarkably good description of minimum bias within a realm of validity that we quantified, performing an explicit comparison to minimum bias collider data. Notably, the (smeared) flat phase space values reproduce kinematic distributions well. We also studied observable features that can be described by an expansion in higher order kinematic harmonics about flat phase space, focusing in particular on azimuthal correlations, specifically elliptic flow in heavy ion collisions and the ridge phenomena in pp collisions. We showed that the conditions of positivity and momentum conservation fix the sign and scaling with $N$ of the leading harmonic of two particle azimuthal difference $\Delta \phi$ beyond the flat limit, and argued that the different symmetries of $e^+e^-$ collisions alone imply additional suppression $1/N^2$ of any ridge that could be observed there. 

The results of the above study are promising, in that they provide confidence that the $S$-matrix framework to describe minimum bias can be developed. Benefits of this approach are that it equally applies to small and large collision system sizes at high and low energies, and it does not depend on any underlying model or unphysical parameters. This could be utilized to elucidate the nature of small scale collective phenomena of QCD that were newly discovered at the LHC, and which are the subject of some debate. In particular, it would be interesting to study the emergence of jets in our picture, and see whether this can address the question of jet quenching in small (potentially QGP phase) systems. Another avenue would be exploring the use of this formalism at  low-energy heavy ion beam scans that aim to hit  the QCD critical point.

Theoretically, there are a number of interesting directions to pursue. A systematic study of the harmonic expansion of the phase space manifold could be undertaken; power spectra of minimum bias collisions can be envisaged from this, and could be an efficient way of determining flow vs.~non-flow physics. It would be interesting and potentially very natural to incorporate detector resolution into this language via, e.g., a maximum harmonic sensitivity.

We formally took a limit $N\to\infty$ while working at fixed $Q$, while ignoring the mass of the QCD pions. This approximation agrees with data well, giving confidence that it is a good approximation to low-energy QCD dynamics in the minimum bias regime. However, strictly this limit does not exist, a maximum multiplicity being ensured by the explicit breaking of chiral symmetry and the pion mass, $N_{\text{max}}=Q/m_{\pi}$. The obvious connection between the limit we worked in and broader studies of strongly-coupled conformal field theories at colliders~\cite{Hofman:2008ar} would be interesting to pursue. It might be possible to address whether  the $Q$ and $N$ dependence of the coefficients the appeared in our expansion be fixed by, e.g., consistency conditions that arise from a non-trivial limit of a distribution as $N\to\infty$.

Finally, by working purely with the  $S$-matrix of final particles, the hydrodynamic description of the unobservable QGP was sidestepped. Nevertheless one might wonder how hallmarks of transport coefficients and, e.g., the theoretical lower bound on specific shear viscosity may show up in our coarse-grained approach. 

\acknowledgments

We thank Bryan Webber for valuable feedback regarding our approach. T.M. is supported by the World Premier International Research Center Initiative (WPI) MEXT, Japan, and by JSPS KAKENHI grants JP19H05810, JP20H01896, and JP20H00153. 

\bibliography{minbias}
\end{document}